# Energy-Efficient Sensor Censoring for Compressive Distributed Sparse Signal Recovery

Jwo-Yuh Wu[*], Ming-Hsun Yang, and Tsang-Yi Wang

***Abstract-*** **To strike a balance between energy efficiency and data quality control, this paper proposes a sensor censoring scheme for distributed sparse signal recovery via compressive-sensing based wireless sensor networks. In the proposed approach, each sensor node employs a sparse sensing vector with known support for data compression, meanwhile enabling making local inference about the unknown support of the sparse signal vector of interest. This naturally leads to a ternary censoring protocol, whereby each sensor (i) directly transmits the real-valued compressed data if the sensing vector support is detected to be overlapped with the signal support, (ii) sends a one-bit hard decision if empty support overlap is inferred, (iii) keeps silent if the measurement is judged to be uninformative. Our design then aims at minimizing the error probability that empty support overlap is decided but otherwise is true, subject to the constraints on a tolerable false-alarm probability that non-empty support overlap is decided but otherwise is true, and a target censoring rate. We derive a closed-form formula of the optimal censoring rule; a low complexity implementation using bi-section search is also developed. In addition, the average communication cost is analyzed. To aid global signal reconstruction under the proposed censoring framework, we propose a modified $\ell_1$-minimization based algorithm, which exploits certain sparse nature of the hard decision vector received at the fusion center. Analytic performance guarantees, characterized in terms of the restricted isometry property, are also derived. Computer simulations are used to illustrate the performance of the proposed scheme.**

# I. INTRODUCTION

*A. Overview*

Compressive sensing (CS) asserts that perfect/stable signal reconstruction from measurements sampled at frequencies far below the Nyquist rate is possible as long as the underlying signal is sparse enough, i.e., it contains only a few nonzeros as compared to the ambient dimension [1-4]. Indeed, CS has been acknowledged as a new signal processing paradigm, in which sparsity is exploited to facilitate efficient signal acquisition and subsequent processing [5-6]. Recently, integration of CS into the design of wireless sensor networks (WSNs) has received considerable attention (among others, [7-15]), mainly because sub-Nyquist sampling economizes the sensor deployment cost. It has been well known that WSNs are subject to limited energy resources [16].

This work is sponsored by the Ministry of Science and Technology of Taiwan under grants MOST 104-2221-E-009-054-MY3 and MOST 105-2221-E-009-011-MY3, and by the Ministry of Education of Taiwan under the MoE ATU Program.

J. Y. Wu and M. H. Yang are with the Department of Electrical and Computer Engineering, and Institute of Communications Engineering, National Chiao Tung University, Taiwan. Emails: jywu@cc.nctu.edu.tw; archenemy310017@hotmail.com.

T. Y. Wang is with the Institute of Communications Engineering, National Sun Yat-sen University, 70, Lien-Hai Road, Kaohsiung, Taiwan. Email: tcwang@faculty.nsysu.edu.tw.

* Contact author.

Among many physical-layer energy conservation schemes tailored for WSNs, *sensor censoring*, whereby only those sensors with good measurement quality are allowed to forward their data to the fusion center (FC), was an effective means for reducing transmission energy [18-20]. In CS-based WSNs, the measurement quality is even more demanding, because the global inference is made on the basis of relatively less data (as compared to the conventional Nyquist based processing). This thus necessitates the development of sensor censoring schemes for CS-based WSNs, in an attempt to conserve transmission energy and, meanwhile, ensure availability of high-quality data to the FC. In-depth study of issues of this kind, however, remains lacking in the literature.

*B. Paper Contribution*

To design sensor censoring schemes for sparse signal recovery via CS-based WSNs, one natural criterion would be minimization of the global signal reconstruction error resulting from a certain signal recovery algorithm, e.g., the $\ell_1$-minimization method [3-4] or greedy-type algorithms such as orthogonal matching pursuit (OMP) [6, Chap. 8]. However, due to the nonlinear nature of CS-based signal reconstruction, it is rather difficult (if not impossible) to analytically characterize the reconstruction error in terms of the censoring system parameters. Hence, rather than adopting parameter estimation based criteria, this paper proposes a *detection theory* oriented censoring scheme. The main idea behind the proposed method is that, at each sensor (say, the *i*th node), a *sparse* sensing vector[1] (with *known* support $\mathcal{A}_i$) is employed to compress the measurement vector, composed of a common desired sparse signal vector (with *unknown* support $\mathcal{T}$) contaminated by noise, into a scalar. Thanks to the sparse nature of sensing vectors, each sensor can make based on the compressed measurement its local inference about the signal support (i.e., a local decision can be made to decide if $\mathcal{T} \cap \mathcal{A}_i$ is empty or not). Only those sensor nodes with reliable decisions are turned on and allowed to forward their censored decisions to the FC. On the basis of this line of thought, specific technical contribution of our paper can be summarized as follows.

1. To exploit the local sensor inference to aid global signal reconstruction, while conserving the transmission energy, a new censoring rule is proposed. Unlike existing binary censoring, i.e., send or no-send schemes, our approach is essentially a ternary protocol: each sensor (i) directly transmits its compressed measurement if $\mathcal{T} \cap \mathcal{A}_i \neq \emptyset$ is decided, (ii) sends a one-bit hard decision when $\mathcal{T} \cap \mathcal{A}_i = \emptyset$ is claimed, and (iii) keeps silent if the measurement is judged to be uninformative. In accordance with such a protocol, the design of our censoring scheme aims at minimizing the error probability of deciding $\mathcal{T} \cap \mathcal{A}_i = \emptyset$ but $\mathcal{T} \cap \mathcal{A}_i \neq \emptyset$ is true, subject to the constraints on (a) a tolerable false-alarm probability of deciding $\mathcal{T} \cap \mathcal{A}_i \neq \emptyset$ while $\mathcal{T} \cap \mathcal{A}_i = \emptyset$ is true, and (b) an allowable censoring rate. We derive a closed-form formula for the optimal censoring rule. A computationally efficient implementation using simple bi-section search is also developed. In addition, the average communication cost per transmission under the proposed censoring protocol is analytically characterized.

---

1. Notably, sparse sensing vectors/matrices have also been considered in the study of CS-based sensing and inference, e.g., [21-22].

2. Toward signal recovery under our censoring protocol, we propose a modified $\ell_1$-minimization based algorithm, which exploits certain sparse nature inherent to the hard decision vector received at the FC. Then, performance guarantees, in terms of the restricted isometry property (RIP) of the sensing matrix, for the proposed signal reconstruction algorithm are further analyzed and derived. Computer simulations confirms that, in the medium-to-high SNR region, the proposed scheme can achieve better signal reconstruction performance using fewer sensor nodes as compared to the conventional CS-based processing without censoring.

The rest of this paper is organized as follows. Section II introduces the system model and the problem statement. Section III presents the proposed censoring scheme. Section IV shows the proposed signal reconstruction algorithm and its performance guarantee. Section V shows the simulation results. Finally, Section VI concludes this paper.

## II. SYSTEM MODEL AND PROBLEM STATEMENT

We consider a WSN, in which $M$ sensor nodes cooperate with a FC for estimating a $K$-sparse signal vector $\mathbf{s} \in \mathbb{R}^N$. Assume that $\mathbf{s}$ is supported on the unknown $\mathcal{T} \subset \{1, \cdots, N\}$ with cardinality $|\mathcal{T}| = K$ ($K \ll N$). For each $1 \leq i \leq M$, the $i$th sensor node first obtains its measurement

$$\mathbf{y}_i = \mathbf{s} + \mathbf{v}_i, \ 1 \leq i \leq M, \tag{2.1}$$

where $\mathbf{v}_i \sim \mathcal{N}(\mathbf{0}_N, \sigma_v^2 \mathbf{I}_N)$ is a zero-mean Gaussian noise vector with covariance $\sigma_v^2 \mathbf{I}_N$, and

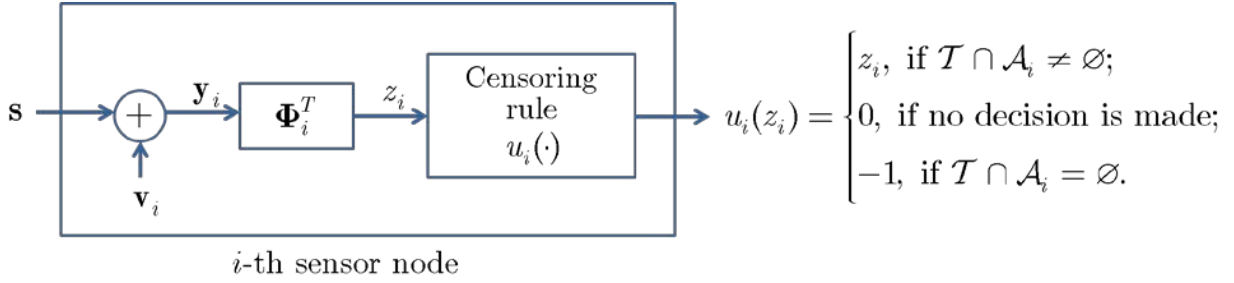

Fig. 1. Depiction of the signal processing protocol of a single sensor node (the $i$-th node).

compresses $\mathbf{y}_i$ into a scalar

$$z_i = \mathbf{\Phi}_i^T \mathbf{y}_i = \mathbf{\Phi}_i^T(\mathbf{s} + \mathbf{v}_i), \ 1 \leq i \leq M, \tag{2.2}$$

where $z_i \in \mathbb{R}$ is the compressed measurement and $\mathbf{\Phi}_i \in \mathbb{R}^N$ is the sensing vector assumed to be $K_c$-sparse with support $\mathcal{A}_i \subset \{1, \cdots, N\}$ ($K_c \geq K$). Thanks to the sparse nature of both the signal $\mathbf{s}$ and sensing vectors $\mathbf{\Phi}_i$'s, (2.2) can be further expressed as

$$z_i = \mathbf{\Phi}_i^T \mathbf{y}_i = \begin{cases} \mathbf{\Phi}_i^T \mathbf{s} + \mathbf{\Phi}_i^T \mathbf{v}_i, & \text{if } \mathcal{T} \cap \mathcal{A}_i \neq \varnothing; \\ \mathbf{\Phi}_i^T \mathbf{v}_i, & \text{if } \mathcal{T} \cap \mathcal{A}_i = \varnothing. \end{cases} \tag{2.3}$$

Notably, when noise is absent, (2.3) is reduced to[2]

$$z_i = \mathbf{\Phi}_i^T \mathbf{y}_i = \begin{cases} \mathbf{\Phi}_i^T \mathbf{s}, & \text{if } \mathcal{T} \cap \mathcal{A}_i \neq \varnothing; \\ 0, & \text{if } \mathcal{T} \cap \mathcal{A}_i = \varnothing, \end{cases} \tag{2.4}$$

which enables us to infer some partial knowledge about the signal support $\mathcal{T}$; for example, certain elements in $\mathcal{A}_i$ shall be included in $\mathcal{T}$ if $z_i \neq 0$, whereas all elements in $\mathcal{A}_i$ can be precluded from $\mathcal{T}$ whenever $z_i = 0$.

To exploit the prior information conveyed by $z_i$ about the unknown signal support when there is noise, we propose to adopt a *ternary censoring rule* $u_i(\cdot)$ at the $i$th node, such that (i) $u_i(z_i) = z_i$ if the $i$th node decides $\mathcal{T} \cap \mathcal{A}_i \neq \varnothing$, (ii) $u_i(z_i) = -1$ if the $i$th node decides[3] $\mathcal{T} \cap \mathcal{A}_i = \varnothing$, and (iii) $u_i(z_i) = 0$ if the $i$th node does not make decision ($z_i$ is deemed to be uninformative); see Fig. 1 for an illustration. Toward a mathematical description of the proposed censoring scheme, for each $1 \leq i \leq M$ let us partition the set of real numbers $\mathbb{R}$ into a disjoint union of three subsets as $\mathbb{R} = \mathcal{R}_0^{(i)} \cup \mathcal{R}_1^{(i)} \cup \mathcal{R}_2^{(i)}$, where

$$\mathcal{R}_0^{(i)} \triangleq \{z_i \in \mathbb{R} \mid u_i(z_i) = -1\} \tag{2.5}$$

and

---

2. We preclude the case that $\mathbf{\Phi}_i^T \mathbf{s} = 0$, if $\mathcal{T} \cap \mathcal{A}_i \neq \varnothing$, which occurs with zero probability.
3. We preclude the case that $z_i = -1$ (and hence $u_i(z_i) = z_i = -1$), if $\mathcal{T} \cap \mathcal{A}_i \neq \varnothing$, which occurs with zero probability.

$$\mathcal{R}_1^{(i)} \triangleq \{z_i \in \mathbb{R} \mid u_i(z_i) = z_i\} \tag{2.6}$$

are the decision regions associated with, respectively, the censored decisions $u_i(z_i) = -1$ and $u_i(z_i) = z_i$, and

$$\mathcal{R}_2^{(i)} \triangleq \{z_i \in \mathbb{R} \mid u_i(z_i) = 0\} \tag{2.7}$$

is the "no-send" region. Notably, the $i$th sensor node is active only when $z_i \in \mathcal{R}_0^{(i)} \cup \mathcal{R}_1^{(i)}$. The design of the censoring rule thus involves determining the three local decision regions, in a way that certain performance requirements are fulfilled. In this paper, we propose a sensor censoring rule for the considered sensing system, to be discussed next. The following assumptions are made in the sequel.

*Assumption 1:* The signal support $\mathcal{T}$ is uniformly drawn from the collection $\mathbf{\Omega}_K \triangleq \{\mathcal{T}_1, \cdots, \mathcal{T}_{C_K^N}\}$ of all $C_K^N = N!/[K!(N-K)!]$ possible sparsity pattern sets, where $\mathcal{T}_j \subset \{1, \cdots, N\}$ with $|\mathcal{T}_j| = K$ and $\Pr\{\mathcal{T}_j\} = 1/C_K^N$. □

*Assumption 2:* The nonzero entries of $\mathbf{s}$, say, $s_k$ for $k \in \mathcal{T}$, are i.i.d. with $s_k \sim \mathcal{N}(0, \sigma_s^2)$, and are independent of the measurement noise $\mathbf{v}_i$'s. □

*Assumption 3:* For each $1 \leq i \leq M$, the sensing vector $\mathbf{\Phi}_i$ is binary with the $K_c$ nonzero entries, i.e., $\mathbf{\Phi}_{ij} \in \{\pm 1\}$, $j \in \mathcal{A}_i$. □

*Assumption 4:* The supports $\mathcal{A}_i, 1 \leq i \leq M$, of the sensing vectors $\mathbf{\Phi}_i$, $1 \leq i \leq M$, are known at the FC. □

Some remarks regarding the proposed sparse-sensing based problem formulation for sensor censoring are in order.

(i) The proposed formulation can be directly applied to the case in which $\mathbf{s}$ admits a sparse representation over some orthonormal basis $\mathbf{B} \in \mathbb{R}^{N \times N}$ other than the identity matrix $\mathbf{I}$, i.e., $\mathbf{s} = \mathbf{B}\mathbf{x}$, where $\mathbf{x} \in \mathbb{R}^N$ is $K$-sparse. Indeed, we choose the sensing vector to be $\mathbf{\Phi}_i = \mathbf{B}\mathbf{\Psi}_i$, where $\mathbf{\Psi}_i \in \mathbb{R}^N$ is $K_c$–sparse. Then the compressed measurement can be expressed as

$$z_i = \mathbf{\Phi}_i^T(\mathbf{s} + \mathbf{v}_i) = \mathbf{\Psi}_i^T \mathbf{B}^T(\mathbf{B}\mathbf{x} + \mathbf{v}_i) \stackrel{(a)}{=} \mathbf{\Psi}_i^T(\mathbf{x} + \mathbf{B}^T \mathbf{v}_i), \ 1 \leq i \leq M, \tag{2.8}$$

where (a) holds since $\mathbf{B}$ is orthonormal, i.e., $\mathbf{B}^T\mathbf{B} = \mathbf{I}$, and $\mathbf{B}^T\mathbf{v}_i \sim \mathcal{N}(\mathbf{0}_N, \sigma_v^2 \mathbf{I}_N)$. Clearly, (2.8) is essentially the same with the sparse sensing measurement model (2.2) (the same noise distribution). Therefore, based on (2.8), the proposed censoring scheme (to be discussed next) can be directly extended to this general case.

(ii) As in many existing CS-based studies of WSNs [23-24], the adopted measurement model (2.1) assumes that each sensor can make a full observation of the signal. Our design framework can also be applied to the scenario in which each sensor only observes a certain part of the signal $\mathbf{s}$. To see this, assume that the $i$th sensor observes the components of $\mathbf{s}$ indexed by $\mathcal{I}_i \subset \{1,\ldots,N\}$. The observation signal $\mathbf{s}_i \in \mathbb{R}^N$ at the $i$th node then reads

$$[\mathbf{s}_i]_j = \begin{cases} [\mathbf{s}]_j, & \text{if } j \in \mathcal{I}_i; \\ 0, & \text{if } j \notin \mathcal{I}_i, \end{cases} \quad (2.9)$$

where $[\mathbf{s}_i]_j$ and $[\mathbf{s}]_j$ are the $j$th elements of $\mathbf{s}_i$ and $\mathbf{s}$, respectively. The compressed measurement can be accordingly expressed as

$$z_i = \mathbf{\Phi}_i^T \mathbf{y}_i = \mathbf{\Phi}_i^T (\mathbf{s}_i + \mathbf{v}_i),\ 1 \leq i \leq M. \quad (2.10)$$

Choose the support of $\mathbf{\Phi}_i$ to be a certain subset of $\mathcal{I}_i$, i.e., $\mathcal{A}_i \subseteq \mathcal{I}_i$. Then we have $\mathbf{\Phi}_i^T \mathbf{s}_i = \mathbf{\Phi}_i^T \mathbf{s}$ and, as a result, the compressed measurement $z_i$ in (2.10) is identical to (2.2). The proposed censoring scheme is thus applicable in this case.

(iii) Even though real-world physical phenomenon and the sensing ranges of practical sensors are bounded, the i.i.d. normal assumption on the signal nonzero entries has been widely considered and adopted in the literature of compressive WSNs [25-27]. The main reasons for this, as far as we can see, are: i) normal random variable/vector is essentially bounded, meaning that its realizations with a very high probability fall within a ball with a sufficiently large but finite radius; ii) normality can greatly facilitate analysis. Notably, the full signal observation model (2.1) together with the Gaussian assumption on the signal nonzero entries has also been considered in [24, 28], which provide a general framework for compressive signal detection for the monitoring application. □

## III. PROPOSED CENSORING SCHEME

This section addresses the design of the censoring scheme at each local node. Section III-A highlights the motivation behind the proposed approach, and presents the problem formulation. Section III-B then derives the optimal censoring rule. Section III-C addresses the computational issues regarding the proposed censoring scheme. Finally, Section III-D discusses the issue of communication cost.

## A. Design Criterion and Problem Formulation

To design the local censoring rule under the CS framework, one conceivable metric from the *parameter estimation* perspective is the global signal reconstruction error achieved by a specifically chosen signal recovery algorithm, e.g., the $\ell_1$-minimization method or greedy-type algorithms such as OMP. However, it is very difficult (if not impossible) to derive an explicit expression of the reconstruction error in terms of the parameters of the local censoring schemes; in particular, how different partitions of $\mathbb{R}$ into $\mathbb{R} = \mathcal{R}_0^{(i)} \cup \mathcal{R}_1^{(i)} \cup \mathcal{R}_2^{(i)}$, for all $1 \leq i \leq M$, can influence the reconstruction error is very hard to characterize. To overcome the above-mentioned difficulties, we propose a censoring design scheme alternatively from the *signal detection* point of view. The proposed approach is mainly based on the following key observations.

- In our setting, the global signal recovery performance at the FC depends crucially on the *reliability of the side information* that can be inferred from the local sensor decisions. Indeed, if for some activated sensor, say, the $j$th node, the true event is $\mathcal{T} \cap \mathcal{A}_j \neq \varnothing$ but an incorrect local decision $\mathcal{T} \cap \mathcal{A}_j = \varnothing$ is made, the FC upon receiving $u_j(z_j) = -1$ will disregard all the indexes in $\mathcal{A}_j$, therefore precluding some elements of $\mathcal{T}$, in the signal reconstruction process: this will cause model mismatch which can largely degrade the signal recovery performance. In light of the above fact, the *local miss-detection probability*, defined to be

$$P_M^{(i)} \triangleq \Pr\{u_i(z_i) = -1 \,|\, \mathcal{T} \cap \mathcal{A}_i \neq \varnothing\}, \tag{3.1}$$

  should be made as small as possible.

- Another type of local detection error is that $\mathcal{T} \cap \mathcal{A}_j = \varnothing$ is true but a decision claims otherwise. Such an event, if it occurs, can be deemed as sort of false-alarm: even though certain indexes are mistaken to belong to the support, the corresponding signal magnitudes actually assume a zero value, making it easy to discard those over-fitted support elements upon signal reconstruction[4]. Over-fitting of the signal support, however, will incur additional data processing complexity. This suggests that the *local false-alarm probability,* defined to be

$$P_F^{(i)} \triangleq \Pr\{u_i(z_i) = z_i \,|\, \mathcal{T} \cap \mathcal{A}_i = \varnothing\}, \tag{3.2}$$

  should be kept small.

---
4. This can be done by, for example, a simple thresholding operation.

- To maintain good global signal reconstruction quality, a sufficient amount of measurements should be made available at the FC. This implies that the *local censoring probability*, that is,

$$P_C^{(i)} \triangleq \Pr\{u_i(z_i) = 0\}, \tag{3.3}$$

cannot be set overly large (otherwise a large portion of sensors will just keep silent).

Motivated by the above facts, a natural criterion for local censoring rule design is to minimize the miss detection probability subject to certain constraints on the false alarm and censoring rates. Specifically, we consider the following optimization problem

$$(P1) \quad \underset{u_i}{Minimize} \, P_M^{(i)}, \text{ subject to } P_C^{(i)} \leq \alpha_i, \, P_F^{(i)} \leq \beta_i,$$

where $\alpha_i$ and $\beta_i$ are the upper bounds of, respectively, the censoring and the local false-alarm probabilities. The optimal solution to Problem $(P1)$ is obtained in the next subsection.

*B. Optimal Censoring Rule*

Toward a solution to $(P1)$, some definitions are needed. Associated with each of the two events $\{\mathcal{T} \cap \mathcal{A}_i = \varnothing\}$ and $\{\mathcal{T} \cap \mathcal{A}_i \neq \varnothing\}$ we define the following two *a priori* probabilities

$$\pi_0^{(i)} \triangleq \Pr(\mathcal{T} \cap \mathcal{A}_i = \varnothing) \text{ and } \pi_1^{(i)} \triangleq \Pr(\mathcal{T} \cap \mathcal{A}_i \neq \varnothing). \tag{3.4}$$

Also, the likelihood ratio (LR) for the measurement $z_i$ is defined to be

$$L(z_i) \triangleq \frac{p(z_i \mid \mathcal{T} \cap \mathcal{A}_i \neq \varnothing)}{p(z_i \mid \mathcal{T} \cap \mathcal{A}_i = \varnothing)}, \tag{3.5}$$

where $p(z_i \mid \mathcal{T} \cap \mathcal{A}_i \neq \varnothing)$ and $p(z_i \mid \mathcal{T} \cap \mathcal{A}_i = \varnothing)$ are the conditional probability density functions of $z_i$. By using standard Lagrange multiplier techniques and exploiting the continuity of $L(\cdot)$, the optimal solution to $(P1)$ is derived in the next theorem.

**Theorem 3.1:** Given the local compressed measurement $z_i$, the optimal censoring rule $u_i^*$ which solves $(P1)$ is given by

$$u_i^*(z_i) = \begin{cases} -1, & \text{if } L(z_i) < \dfrac{\lambda_2^{(i)} \pi_0^{(i)}}{1 - \lambda_2^{(i)} \pi_1^{(i)}}; \\ z_i, & \text{if } L(z_i) > \dfrac{\lambda_1^{(i)} - \lambda_2^{(i)} \pi_0^{(i)}}{\lambda_2^{(i)} \pi_1^{(i)}}; \\ 0, & \text{if } \dfrac{\lambda_2^{(i)} \pi_0^{(i)}}{1 - \lambda_2^{(i)} \pi_1^{(i)}} < L(z_i) < \dfrac{\lambda_1^{(i)} - \lambda_2^{(i)} \pi_0^{(i)}}{\lambda_2^{(i)} \pi_1^{(i)}}, \end{cases} \tag{3.6}$$

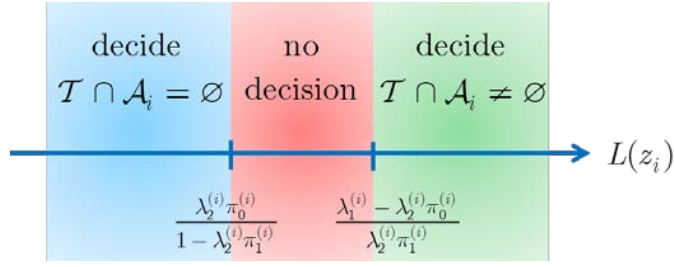

Fig. 2. Depiction of the optimal censoring rule using the LR.

where $\pi_0^{(i)}$ and $\pi_1^{(i)}$ are defined in (3.4), $L(z_i)$ is defined in (3.5), and $\lambda_1^{(i)}$ and $\lambda_2^{(i)}$ are the Lagrange multipliers chosen so that the constraints in ($P$1) are satisfied.

*[Proof]:* See Appendix A. □

Note that the optimal censoring rule (3.6) first computes the LR $L(z_i)$, and then compares it to a certain threshold to make the local decision (see Fig. 2 for a depiction); also note that (3.6) holds true irrespective of the probability distribution[5] of $z_i$. By invoking Assumptions 1, 2, and 3, the formula for the optimal censoring rule (3.6) can be simplified even further. For this we first derive closed-form formulae for $p(z_i \mid \mathcal{T} \cap \mathcal{A}_i \neq \varnothing)$ and $p(z_i \mid \mathcal{T} \cap \mathcal{A}_i = \varnothing)$, as given in the next lemma.

**Lemma 3.2:** For each $1 \leq i \leq M$, the conditional probability density functions $p(z_i \mid \mathcal{T} \cap \mathcal{A}_i \neq \varnothing)$ and $p(z_i \mid \mathcal{T} \cap \mathcal{A}_i = \varnothing)$ can be respectively expressed as

$$p(z_i \mid \mathcal{T} \cap \mathcal{A}_i \neq \varnothing) = \sum_{j=1}^{K} P_j \frac{1}{\sqrt{2\pi\left(j\sigma_s^2 + K_c\sigma_v^2\right)}} \exp\left(-\frac{z_i^2}{2\left(j\sigma_s^2 + K_c\sigma_v^2\right)}\right), \quad (3.7)$$

where

$$P_j \triangleq \Pr\{|\mathcal{T} \cap \mathcal{A}_i| = j \mid \mathcal{T} \cap \mathcal{A}_i \neq \varnothing\} = C_j^{K_c} C_{K-j}^{N-K_c} / \left(\sum_{j'=1}^{K} C_{j'}^{K_c} C_{K-j'}^{N-K_c}\right) \quad (3.8)$$

is the probability that the cardinality of $\mathcal{T} \cap \mathcal{A}_i$ is $j$ given that $\mathcal{T} \cap \mathcal{A}_i$ is nonempty, and

$$p(z_i \mid \mathcal{T} \cap \mathcal{A}_i = \varnothing) = \frac{1}{\sqrt{2\pi K_c \sigma_v^2}} \exp\left(-\frac{z_i^2}{2K_c\sigma_v^2}\right). \quad (3.9)$$

*[Proof]:* See Appendix B. □

By means of (3.7), (3.9), and with some manipulations, the LR in (3.5) can be expressed as

$$L(z_i) = \sum_{j=1}^{K} P_j \sqrt{\frac{K_c \sigma_v^2}{j\sigma_s^2 + K_c \sigma_v^2}} \exp\left(\frac{j\sigma_s^2 z_i^2}{2K_c \sigma_v^2 (j\sigma_s^2 + K_c \sigma_v^2)}\right). \tag{3.10}$$

---

5. Thanks to this nice feature and based on Remark (ii) at the end of Section II, the proposed approach can be directly applied to the partial observation model (2.9)-(2.10) with non-Gaussian signal distribution. The Gaussianality signal assumption, however, does facilitate efficient computation of the optimal censoring rule, as will be seen in the rest of this section.

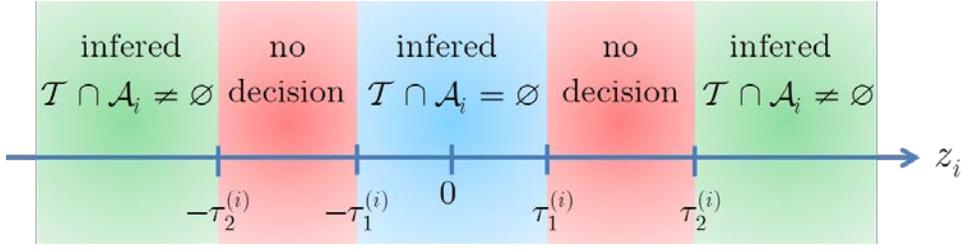

Fig. 3. Depiction of the optimal censoring rule directly in the measurement domain.

Using (3.10), the next lemma further characterizes a certain restriction of the function $L(\cdot)$ that will be used in the follow-up discussions.

***Lemma 3.3:*** Let $L_0(\cdot): \mathbb{R}^+ \cup \{0\} \to \mathbb{R}^+ \cup \{0\}$ be the restriction of $L(\cdot)$ on $\mathbb{R}^+ \cup \{0\}$, that is, $L_0 = L \mid_{\mathbb{R}^+ \cup \{0\}}$. Then $L_0(\cdot)$ is monotonically increasing, and is thus one-to-one.

*[Proof]:* See Appendix C. □

Based on Lemmas 3.2 and 3.3, the optimal censoring rule $u_i^*$ in (3.6) can be shown to admit a very simple form, as established the next theorem.

***Theorem 3.4:*** Let $L_0(\cdot)$ be defined as in Lemma 3.3, and $L_0^{-1}(\cdot)$ be the corresponding inverse function. The optimal censoring rule $u_i^*$ in (3.6) can be expressed as

$$u_i^*(z_i) = \begin{cases} -1, & \text{if } |z_i| < \tau_1^{(i)}; \\ z_i, & \text{if } |z_i| > \tau_2^{(i)}; \\ 0, & \text{if } \tau_1^{(i)} < |z_i| < \tau_2^{(i)}, \end{cases} \tag{3.11}$$

where

$$\tau_1^{(i)} \triangleq L_0^{-1}\left(\frac{\lambda_2^{(i)} \pi_0}{1 - \lambda_2^{(i)} \pi_1}\right), \quad \tau_2^{(i)} \triangleq L_0^{-1}\left(\frac{\lambda_1^{(i)} - \lambda_2^{(i)} \pi_0}{\lambda_2^{(i)} \pi_1}\right), \tag{3.12}$$

and

$$\pi_0 \triangleq C_K^{N-K_c}/C_K^N, \quad \pi_1 \triangleq \sum_{j=1}^{K} C_j^{K_c} C_{K-j}^{N-K_c}/C_K^N. \tag{3.13}$$

*[Proof]:* See Appendix D. □

As the theorem shows, the optimal censoring rule just compares the measurement amplitude $|z_i|$ to certain thresholds as in (3.11), without the need of computing the LR $L(z_i)$ (see also Fig. 3 for a depiction). Notably, implementation of the proposed scheme requires the computation of the two thresholds $\tau_1^{(i)}$ and $\tau_2^{(i)}$ in (3.12), which involve the knowledge of the inverse function $L_0^{-1}(\cdot)$ and the pair of Lagrange multipliers $\lambda_1^{(i)}$ and $\lambda_2^{(i)}$. However, closed-form expressions for $L_0^{-1}(\cdot)$ and $(\lambda_1^{(i)}, \lambda_2^{(i)})$ are rather hard to find. To resolve this difficulty, below we thus propose a method for efficient implementation of the optimal censoring rule (3.11).

*C. Implementation of the Optimal Censoring Scheme*

To proceed, we first define the function $g: \mathbb{R} \to \mathbb{R}$ as

$$g(x) = 2\pi_0 Q\left(\frac{x}{\sqrt{K_c}\sigma_v}\right) + 2\pi_1 \sum_{j=1}^{K} P_j Q\left(\frac{x}{\sqrt{j\sigma_s^2 + K_c \sigma_v^2}}\right), \quad (3.14)$$

$Q$ is the standard $Q$-function. Clearly, $g$ is monotonically decreasing and surjective; thus, the corresponding inverse function, denoted by $g^{-1}$, exists. Using (3.14), formulas of the two thresholds $\tau_1^{(i)}$ and $\tau_2^{(i)}$ in (3.12) are given in the following theorem.

***Theorem 3.5:*** Let $g(\cdot)$ be defined as in (3.14) and $g^{-1}(\cdot)$ be the corresponding inverse function. Then we have

$$\tau_1^{(i)^*} = g^{-1}\left(\alpha_i + \pi_0 \beta_i + 2\pi_1 \sum_{j=1}^{K} P_j Q\left(\frac{\sqrt{K_c}\sigma_v Q^{-1}\left(\frac{\beta_i}{2}\right)}{\sqrt{j\sigma_s^2 + K_c\sigma_v^2}}\right)\right), \quad (3.15)$$

and

$$\tau_2^{(i)^*} = \sqrt{K_c}\sigma_v Q^{-1}\left(\frac{\beta_i}{2}\right), \quad (3.16)$$

where $P_j$ is defined in (3.8), and $\alpha_i$ and $\beta_i$ are defined right after (P1).

*[Proof]:* See Appendix E. □

Theorem 3.5 asserts that, to obtain the optimal thresholds, neither do we need to determine the inverse function $L_0^{-1}(\cdot)$ nor compute the Lagrange multipliers $\lambda_1^{(i)}$ and $\lambda_2^{(i)}$. In addition, with the aid of (3.14) and (3.15), the optimal thresholds can be obtained by using numerical search, such as the bi-section method [29].

*D. Issues of Communication Cost*

In our setting, the *local expected communication cost* is given by

$$C_i \triangleq C_{0i} \Pr\{z_i \in \mathcal{R}_0^{(i)}\} + C_{1i} \Pr\{z_i \in \mathcal{R}_1^{(i)}\}, \quad (3.17)$$

where $C_{0i} > 0$ and $C_{1i} > 0$ represent the costs per transmission when the $i$th node transmits,

respectively, the hard decision $-1$ and the real-valued measurement $z_i$. In general, transmitting a real-valued data requires a higher communication cost[6]; thus, we assume $C_{1i} > C_{0i}$ for all $i$. Based on Theorems 3.4 and 3.5, $C_i$ in (3.17) admits a simple closed-form expression, as shown in the following theorem.

**Theorem 3.6:** For each $1 \leq i \leq M$, the expected communication cost $C_i$ can be expressed as

$$C_i = C_{0i}(1-\alpha_i) + (C_{1i} - C_{0i})g\left[\sqrt{K_c}\sigma_v Q^{-1}\left(\frac{\beta_i}{2}\right)\right], \tag{3.18}$$

where $g$ is defined in (3.14), $\alpha_i$ and $\beta_i$ are defined right after ($P$1).

*[Proof]:* See Appendix F. □

The closed-form formula (3.18) for the mean communication cost $C_i$ affords interesting interpretations as discussed below.

(1) Since $C_{0i} > 0$, the cost $C_i$ decreases as the value of $\alpha_i$ increases. To see this, recall that $\alpha_i$ is the upper bound of no-send probability at $i$th node. Hence, a censoring rule based on a large value of $\alpha_i$ tends to shut down a large portion of sensors, resulting in the low average transmission cost.

(2) Since both the functions $g$ and $Q$ are monotonically decreasing, the composite function $g \circ \sqrt{K_c}\sigma_v Q^{-1}$ is monotonically increasing; this implies that the expected communication cost $C_i$ decreases as the value of $\beta_i$ is decreased. The reason is that $\beta_i$ reflects the frequency of false-alarm occurrence. Hence, a small value of $\beta_i$ implies the active nodes will transmit censored decisions $u_i(z_i) = z_i$ with low probability. Therefore, as real-valued data are transmitted less frequent, the corresponding average communication cost is reduced.

(3) Finally we note that, for a given allowable communication cost $C_i$, we can utilize (3.18) to determine the corresponding $\alpha_i$ and $\beta_i$, under which the expected communication introduced by the optimal censoring rule (3.11) can meet the target cost $C_i$.

---

6. In digital communications, high transmission quality of real-valued data can be achieved by using multiple-bit quantization which can afford high resolution [30-31]. Hence, as compared with 1-bit data transmission, it requires more communication cost.

# IV. SIGNAL RECONSTRUCTION

This section studies the issue of signal reconstruction under the proposed censoring protocol. Section IV-A presents the receive signal model at the FC. Section IV-B introduces the proposed signal reconstruction algorithm. Section IV-C then derives the performance guarantees.

*A. Signal Model*

To reduce network-wide energy consumption, only those sensors with measurements $z_i$'s judged to be informative will forward their censored decisions (i.e., $u_i(z_i)=z_i$ or $u_i(z_i)=-1$) to the FC for global signal reconstruction; nodes with $u_i(z_i)=0$ just keep silent[7]. Notably, an active node $i$ with $u_i(z_i)=-1$ decides $\mathcal{T} \cap \mathcal{A}_i = \varnothing$; if this decision is correct, we have $\mathbf{\Phi}_i^T \mathbf{s} = 0$, i.e., the sensing operation nulls the signal, leaving the compressed measurement $z_i$ in (2.3) composed of noise only. To exploit censoring information of this kind for signal reconstruction, we will replace each received hard decision outcome $u_i(z_i)=-1$ by the corresponding noiseless measurement $u_i(z_i) = \mathbf{\Phi}_i^T \mathbf{s} = 0$; this amounts to "cleaning up" the noisy data (as the noise is independent of the signal).

In the sequel, we define the index subset of all active nodes to be

$$\mathcal{S} \triangleq \mathcal{I} \cup \mathcal{I}_{-1} \subseteq \{1, \cdots, M\}, \qquad (4.1)$$

where

$$\mathcal{I} \triangleq \{i \mid z_i \in \mathcal{R}_1^{(i)}\} \text{ and } \mathcal{I}_{-1} \triangleq \{i \mid z_i \in \mathcal{R}_0^{(i)}\}. \qquad (4.2)$$

Thus, the set of censored decisions received at the FC is

$$\{u_i(z_i)\}_{i \in \mathcal{S}} = \{u_i(z_i)\}_{i \in \mathcal{I}} \cup \{u_i(z_i)\}_{i \in \mathcal{I}_{-1}} = \{z_i\}_{i \in \mathcal{I}} \cup \{0\}_{i \in \mathcal{I}_{-1}}. \qquad (4.3)$$

Toward a mathematical model for signal reconstruction, we stack all $\mathbf{\Phi}_i$'s to get

$$\mathbf{\Phi} \triangleq [\mathbf{\Phi}_1 \cdots \mathbf{\Phi}_M]^T \in \mathbb{R}^{M \times N}. \qquad (4.4)$$

Collecting all the received data in (4.3) into a vector, we obtain[8]

$$\begin{bmatrix} \mathbf{u}_{\mathcal{I}} \\ \mathbf{u}_{\mathcal{I}_{-1}} \end{bmatrix} = \begin{bmatrix} \mathbf{\Phi}_{\mathcal{I}} \\ \mathbf{\Phi}_{\mathcal{I}_{-1}} \end{bmatrix} \mathbf{s} + \mathbf{w}, \qquad (4.5)$$

---

7. The proposed ternary censoring protocol relies on the perfect synchronization assumption, which has been widely considered and adopted in the literature of WSNs [32-34].

8. As in many existing works, e.g., [35-37], our study assumes perfect data reception at the FC. Design of censoring schemes further taking into account the communication link errors (e.g., under Rayleigh or multiple-access interference as in [11]) is our future work.

where $\mathbf{u}_{\mathcal{I}} \in \mathbb{R}^{|\mathcal{I}|}$ consists of $\{z_i\}_{i \in \mathcal{I}}$, $\mathbf{u}_{\mathcal{I}_{-1}} \in \mathbb{R}^{|\mathcal{I}_{-1}|}$ is a vector of all zeros, $\mathbf{\Phi}_{\mathcal{I}} \in \mathbb{R}^{|\mathcal{I}| \times N}$ and $\mathbf{\Phi}_{\mathcal{I}_{-1}} \in \mathbb{R}^{|\mathcal{I}_{-1}| \times N}$ are obtained by retaining the rows of the sensing matrix $\mathbf{\Phi}$ indexed by $\mathcal{I}$ and $\mathcal{I}_{-1}$, respectively, and $\mathbf{w} \in \mathbb{R}^{|\mathcal{S}|}$ is the data mismatch term caused by the sensing noise and local decision errors[9]. With (4.5), an $\ell_1$-minimization based signal reconstruction algorithm tailored for the proposed censoring scheme is developed in the next subsection.

## B. Proposed Signal Reconstruction Scheme

To see the motivation behind our approach, we recall that, when all the nodes indexed by $\mathcal{I}_{-1}$ make correct decisions, the compressed signal $\mathbf{\Phi}_{\mathcal{I}_{-1}}\mathbf{s}$ will be identically zero. In the high-SNR regime, it is expected that those sensors will make erroneous decisions with a very small probability; accordingly, $\mathbf{\Phi}_{\mathcal{I}_{-1}}\mathbf{s}$ will contain only a few non-zeros, rendering $\mathbf{\Phi}_{\mathcal{I}_{-1}}\mathbf{s}$ a sparse vector. Hence, in addition to promoting the sparsity of the signal $\mathbf{s}$, we shall also exploit the sparse nature of $\mathbf{\Phi}_{\mathcal{I}_{-1}}\mathbf{s}$ to facilitate signal reconstruction. Specifically, we propose to solve the following optimization problem for signal recovery

$$(P2) \quad \underset{\mathbf{s}}{Minimize} \|\mathbf{s}\|_1 + \lambda \|\mathbf{\Phi}_{\mathcal{I}_{-1}}\mathbf{s}\|_1, \text{ subject to } \|\mathbf{u}_{\mathcal{I}} - \mathbf{\Phi}_{\mathcal{I}}\mathbf{s}\|_2 \leq \varepsilon$$

In $(P2)$, $\lambda > 0$ is the weighting parameter and $\varepsilon > 0$ specifies the error level caused by sensing noise. Note that the $\ell_1$-norm based objective function in $(P2)$ can be rewritten as

$$\|\mathbf{s}\|_1 + \lambda \|\mathbf{\Phi}_{\mathcal{I}_{-1}}\mathbf{s}\|_1 = \|\mathbf{A}\mathbf{s}\|_1, \text{ where } \mathbf{A} \triangleq [\mathbf{I} \quad \lambda \mathbf{\Phi}_{\mathcal{I}_{-1}}^T]^T \in \mathbb{R}^{(N + |\mathcal{I}_{-1}|) \times N}. \tag{4.6}$$

Hence, $(P2)$ is equivalent to the following weighted $\ell_1$-norm minimization problem

$$(P3) \quad \underset{\mathbf{s}}{Minimize} \|\mathbf{A}\mathbf{s}\|_1, \text{ subject to } \|\mathbf{u}_{\mathcal{I}} - \mathbf{\Phi}_{\mathcal{I}}\mathbf{s}\|_2 \leq \varepsilon,$$

which is a standard convex optimization problem, and can be solved by using numerically efficient tools.

*Remarks:*

(i) To solve $(P3)$, it is implicitly assumed that the data mismatch is bounded, and knowledge of the error level $\varepsilon$ is required. In our signal model (4.5), the sensing noise is Gaussian (cf. (2.1)), which is unbounded. However, Gaussian noise is commonly deemed to be "essentially bounded," meaning that the noise amplitudes most of the time fall below a sufficiently large threshold. In our simulations, the error level $\varepsilon$ is simply set to be the average noise standard deviation, equal to $\varepsilon = \sigma_v \sqrt{K_c |\mathcal{I}|}$; this setting is seen to yield a satisfactory global signal reconstruction performance. We note that $\ell_1$-minimization based methods with a bounded $\ell_2$-norm error constraint have been widely used for CS based studies of communication systems corrupted with Gaussian channel noise [38-42], wherein the bound in the $\ell_2$-norm error constraint is also set to be a scaled noise standard deviation.

---

9. If we accordingly partition $\mathbf{w}$ into $\mathbf{w} = [\mathbf{w}_{\mathcal{I}}^T \, \mathbf{w}_{\mathcal{I}_{-1}}^T]^T$, $\mathbf{w}_{\mathcal{I}}$ accounts for the sensing noise, and $\mathbf{w}_{\mathcal{I}_{-1}}$ is due to local decision errors.

(ii) To solve the convex optimization problem $(P3)$, once can adopt the interior point algorithm [43], which results in algorithmic complexity of order $O(|\mathcal{I}|^2 (N + |\mathcal{I}_{-1}|)^{\frac{3}{2}})$. The corresponding complexity of solving the conventional $\ell_1$-minimization scheme is on the order of $O((|\mathcal{I}| + |\mathcal{I}_{-1}|)^2 N^{\frac{3}{2}})$. When $|\mathcal{I}|/M = O(1)$ and $|\mathcal{I}_{-1}|/M = O(1)$, the algorithmic complexities of the proposed method and the conventional $\ell_1$-minimization scheme are of orders $O(M^2(N+M)^{\frac{3}{2}})$ and $O(M^2(N)^{\frac{3}{2}})$, respectively. As expected, our method requires more computations, since the sparse nature of $\Phi_{\mathcal{I}_{-1}}\mathbf{s}$ is further exploited to facilitate signal reconstruction. □

## C. RIP-based Performance Guarantees

The RIP-based performance guarantee of the proposed signal reconstruction scheme under Gaussian sensing noise is given in the following theorem.

***Theorem 4.1:*** Let $\Phi_{\mathcal{I}}$ and $\mathbf{A}$ be defined in (4.5) and (4.6), respectively, and $\sigma_K(\mathbf{As})_1 \triangleq \min_{\|\mathbf{v}\|_0 \leq K} \|\mathbf{As} - \mathbf{v}\|_1$. If the matrix $\frac{1}{|\mathcal{I}|}\Phi_{\mathcal{I}}\mathbf{A}^\dagger$, where $\mathbf{A}^\dagger$ is the pseudo-inverse of $\mathbf{A}$, satisfies the RIP of order $2K$ with $\delta_{2K} < \sqrt{2} - 1$, the solution $\hat{\mathbf{s}}$ to $(P3)$ satisfies

$$\|\hat{\mathbf{s}} - \mathbf{s}\|_2 \leq \frac{2\left[1 - (1-\sqrt{2})\delta_{2K}\right]}{\left[1 - (1+\sqrt{2})\delta_{2K}\right]\sigma_{\min}(\mathbf{A})} \frac{\sigma_K(\mathbf{As})_1}{\sqrt{K}} + \frac{4\sqrt{1+\delta_{2K}}}{\left[1 - (1+\sqrt{2})\delta_{2K}\right]\sigma_{\min}(\mathbf{A})} \frac{\varepsilon}{|\mathcal{I}|} \quad (4.7)$$

with probability at least $1 - \exp(-c_3 |\mathcal{I}|\varepsilon')$, where $c_3 > 0$ is a constant, $\varepsilon' = \left(\frac{\varepsilon}{\sqrt{K_c |\mathcal{I}|}\sigma_v} - 1\right)^2$ and $\sigma_{\min}(\mathbf{A}) \geq 1$ is the smallest singular value of $\mathbf{A}$.

*[Proof]:* See Appendix G. □

Due to Gaussian sensing noise, the reconstruction error upper bound (4.7) holds under a certain probability; notably, performance guarantees in a similar probability sense under Gaussian noise corruption has been seen in the literature of CS, e.g., [44-45]. Theorem 4.1 relies on the assumption that the matrix $\frac{1}{|\mathcal{I}|}\mathbf{\Phi}_\mathcal{I}\mathbf{A}^\dagger$ satisfies RIP with a small constant. When resorting to random sensing [1-6] commonly considered in CS, it can be shown that $\frac{1}{|\mathcal{I}|}\mathbf{\Phi}_\mathcal{I}\mathbf{A}^\dagger$ satisfies RIP with an overwhelming probability if the binary non-zero entries of $\mathbf{\Phi}_i$'s are populated according to certain distributions. To proceed, we first make the following two assumptions.

***Assumption 5:*** Assume that the non-zeros of $\mathbf{\Phi}_i$ are independently drawn from the symmetric Bernoulli distribution, i.e., $\mathbf{\Phi}_{ij} \in \{\pm 1\}$ with $\Pr\{\mathbf{\Phi}_{ij} = 1\} = \Pr\{\mathbf{\Phi}_{ij} = -1\} = 1/2$ for $j \in \mathcal{A}_i$. □

***Assumption 6:*** For each $1 \leq i \leq M$, the sensing vector support $\mathcal{A}_i$ is uniformly drawn from the collection $\mathbf{\Omega}_{K_c} \triangleq \{\mathcal{A}_{i,1}, \cdots, \mathcal{A}_{i,C_{K_c}^N}\}$ of all $C_{K_c}^N = N!/[K_c!(N-K_c)!]$ possible sparsity pattern sets, where $\mathcal{A}_{i,j} \subset \{1, \cdots, N\}$ with $|\mathcal{A}_{i,j}| = K_c$ and $\Pr\{\mathcal{A}_{i,j}\} = 1/C_{K_c}^N$. □

Under Assumptions 5 and 6, we have the following.

***Theorem 4.2:*** Let $\mathbf{\Phi}_\mathcal{I}$ and $\mathbf{A}$ be defined in (4.5) and (4.6), respectively. Also, let $s_{\min}(\mathbf{A}^\dagger, K) \triangleq \min_{\|\mathbf{v}\|_0 \leq K, \|\mathbf{v}\|_2 = 1} \|\mathbf{A}^\dagger \mathbf{v}\|_2$ and $s_{\max}(\mathbf{A}^\dagger, K) \triangleq \max_{\|\mathbf{v}\|_0 \leq K, \|\mathbf{v}\|_2 = 1} \|\mathbf{A}^\dagger \mathbf{v}\|_2$. Then, for every sparsity level $1 \leq K \leq \lfloor (|\mathcal{I}_{-1}| + N)/2 \rfloor$ and every $\delta_K \in (1 - \rho s_{\min}^2(\mathbf{A}^\dagger, K), 1)$, where $\rho \triangleq K_c/N$, if

$$|\mathcal{I}| > \frac{c_1 K s_{\max}^2(\mathbf{A}^\dagger, K)}{\rho^2 \theta^2(\delta_K) s_{\min}^2(\mathbf{A}^\dagger, K)} \log\left(\frac{5e(|\mathcal{I}_{-1}| + N)}{K}\right), \quad (4.8)$$

the matrix $\frac{1}{|\mathcal{I}|}\mathbf{\Phi}_\mathcal{I}\mathbf{A}^\dagger$ satisfies the RIP of order $K$ with probability exceeding $1 - \exp(-c_2 \rho^2 \theta^2(\delta_K)|\mathcal{I}|)$, where $c_1$ and $c_2$ are positive absolute constants and $0 < \theta(\delta_K) < 1$ is defined as

$$\theta(\delta_K) = 1 - \left(\frac{1 - \delta_K}{\rho s_{\min}^2(\mathbf{A}^\dagger, K)}\right). \quad (4.9)$$

*[Proof]:* See Appendix H. □

## V. SIMULATION RESULTS

In this section, computer simulations are used to illustrate the performance of the proposed scheme. The ambient signal dimension is set to be $N = 500$; the sparsity level of the desired signal **s** is $K = 5$, and the number of non-zeros in the compression vectors is $K_c = 20$. We consider a homogeneous sensing environment, wherein the censoring probabilities $\alpha_i$'s and the false alarm rates $\beta_i$'s are identical across all sensor nodes, that is, $\alpha_i = \alpha$ and $\beta_i = \beta$, for all $1 \leq i \leq M$. The SNR of the local sensor measurement is defined as $\text{SNR} \triangleq E\{\|\mathbf{s}\|^2\}/E\{\|\mathbf{v}_i\|^2\}$. The quality of signal recovery is evaluated by using the normalized mean square error (NMSE), defined as [11]

$$NMSE \triangleq E\{\|\mathbf{s} - \hat{\mathbf{s}}\|^2 / \|\mathbf{s}\|^2\}, \quad (5.1)$$

where $\hat{\mathbf{s}}$ is the reconstructed sparse signal at the FC. To assess the achieved energy efficiency of our censoring policy, the fraction of active nodes (FAN) is used as the metric:

$$\text{FAN} \triangleq \frac{\text{number of active nodes}}{\text{total number of sensors}}. \quad (5.2)$$

The following three signal processing protocols will be considered in our simulation:

1. Conventional CS method, which activates all sensor nodes and utilizes standard $\ell_1$-minimization [4] for signal reconstruction (abb. as CS-$\ell_1$);
2. CS in conjunction with the proposed censoring scheme, using standard $\ell_1$-minimization [4] for signal reconstruction (abb. as CSC-$\ell_1$);
3. CS in conjunction with the proposed sensor censoring scheme, using the modified $\ell_1$-minimization method (by solving $(P2)$) for signal recovery (abb. CSC-modified_$\ell_1$).

*A. Reconstruction Performance versus Network Size*

We first compare the performances of the three methods with different network size (*M*). For $\alpha = 0.5$, $\beta = 0.075$ and SNR=9 dB, Figs. 4 and 5 plot the NMSE and the corresponding FAN, respectively, for the three methods mentioned above. The results clearly demonstrate the effectiveness of the proposed censoring scheme: even activating only half of the total sensor nodes, CSC-$\ell_1$ performs nearly as well as CS-$\ell_1$, whereas CSC-modified_$\ell_1$ achieves a further lower NMSE. Hence, under a reasonably good SNR condition, just turning on those sensors with good local measurement quality (as judged by the proposed censoring rule) can improve the global signal reconstruction performance. The reason CSC-modified_$\ell_1$ can outperform CSC-$\ell_1$ is that, in the medium-to-high SNR regime, the local sensor decisions (in particular, the hard decisions $u_i(z_i) = -1$ which claims $\mathcal{T} \cap \mathcal{A}_i = \varnothing$) are mostly correct, thereby rendering the vector $\mathbf{\Phi}_{\mathcal{I}_{-1}}\mathbf{s}$ likely to be sparse. As a result, the proposed modified-$\ell_1$ method $(P2)$ performs better than the standard $\ell_1$-minimization, because of promoting the sparsity of $\mathbf{\Phi}_{\mathcal{I}_{-1}}\mathbf{s}$.

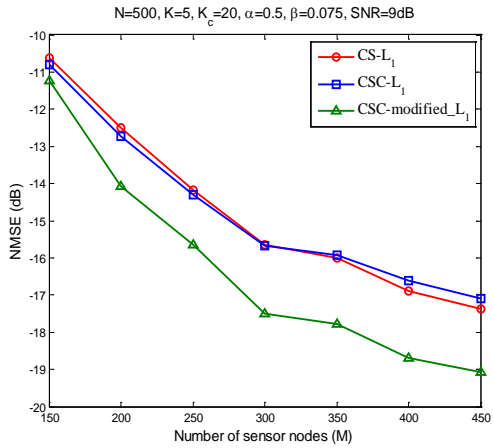
Fig. 4. NMSE of three signal reconstruction methods as a function of $M$.

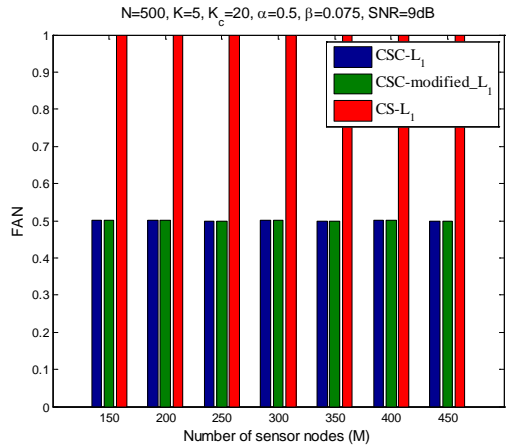
Fig. 5. FAN of three signal reconstruction methods as a function of $M$.

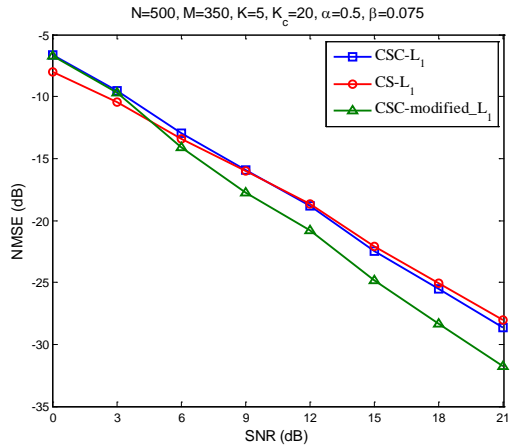
Fig. 6. NMSE of three signal reconstruction methods as a function of SNR.

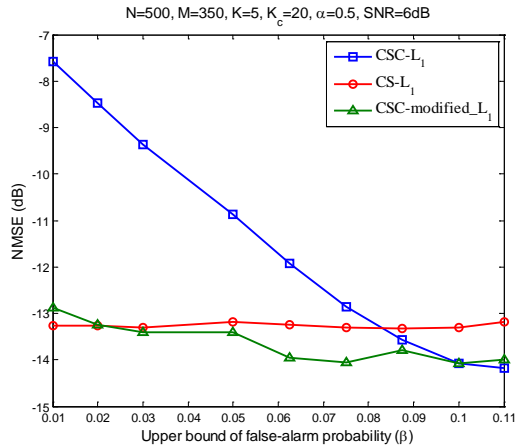
Fig. 7. NMSE of three signal reconstruction methods as a function of $\beta$.

*B. Reconstruction Performance versus Local Sensing SNR*

With a network size $M = 350$, Fig. 6 plots the NMSE for the three methods with respect to different values of sensing SNR. As the figure shows, CSC-modified_$\ell_1$ achieves the lowest NMSE in the medium-to-high SNR region (above 5 dB), as has been explained in Section V-A. Notably, CS-$\ell_1$ performs better than CSC-$\ell_1$ and CSC-modified_$\ell_1$ when SNR is low. This is not unexpected since CS-$\ell_1$ activates all sensor nodes and, thus, uses all real-valued measurements for signal reconstruction: this in turn achieves better noise mitigation and reduces NMSE.

*C. Effect of False-Alarm Probability $\beta$*

For $M = 350$, $\alpha = 0.5$, and SNR=6 dB, Fig. 7 plots the NMSE of the three methods with respect to different values of the false-alarm probability $\beta$. We first note from the figure that, since CS-$\ell_1$ does not employ sensor censoring, the achieved NMSE is a constant (equal to about $-13.3$ dB) irrespective of the values of $\beta$. Also, CSC-modified_$\ell_1$ achieves the lowest NMSE for $0.025 \leq \beta \leq 0.1$. If $\beta$ is large ($\beta > 0.1$), more active sensor nodes will forward real-valued measurements $u_i(z_i) = z_i$, rather than the hard decision $u_i(z_i) = -1$, to the FC. In this case, CSC-$\ell_1$ outperforms CSC-modified_$\ell_1$ since the dimension of $\Phi_{\mathcal{I}_{-1}}\mathbf{s}$ is reduced and promoting sparsity of $\Phi_{\mathcal{I}_{-1}}\mathbf{s}$ does not afford substantial performance gain. On the other hand, if $\beta$ is small ($\beta < 0.025$), most of the active nodes will send the hard decisions $u_i(z_i) = -1$. As such, only a few real-valued measurements are available for signal reconstruction when sensor censoring is employed; instead, CS-$\ell_1$ has access to all real-valued sensor measurements and can achieve the best performance.

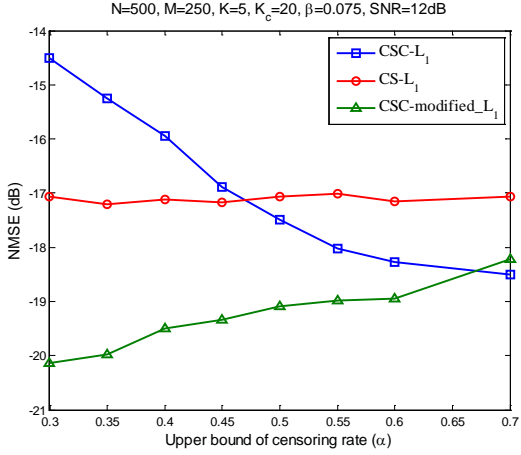 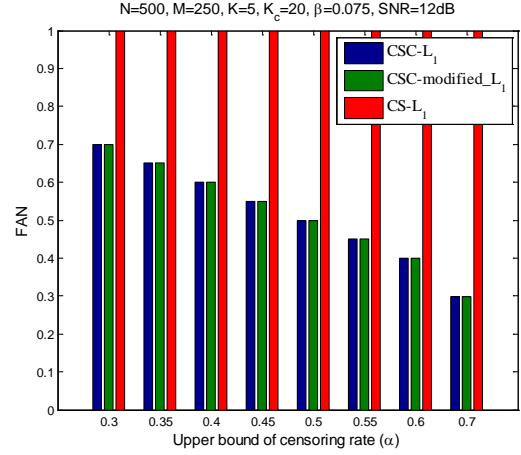

Fig. 8. NMSE of three signal reconstruction methods as a function $\alpha$.

Fig. 9. FAN of three signal reconstruction methods as a function $\alpha$.

## D. Effect of Censoring Rate $\alpha$

With $M = 250$, $\beta = 0.075$ and SNR=12 dB, Fig. 8 depicts NMSE versus the upper bounds of censoring rate $\alpha$. Still, CS-$\ell_1$ attains a constant NMSE (equal to about $-17.1$ dB) irrespective of the values of $\alpha$. The figure shows that, for small $\alpha$, CSC-modified_$\ell_1$ achieves the lowest NMSE. This is because, if $\alpha$ is small, more sensors will be activated and, in particular, the number of nodes transmitting $u_i(z_i) = -1$ increases. In this way, $\Phi_{\mathcal{I}_{-1}}\mathbf{s}$ is more likely to be sparse, and CSC-modified_$\ell_1$ (which promotes sparsity of $\Phi_{\mathcal{I}_{-1}}\mathbf{s}$) yields improved performance. As $\alpha$ grows, say, above $0.65$, a large portion of sensors is kept silent; as such, $\Phi_{\mathcal{I}_{-1}}\mathbf{s}$ is of a small dimension or rendered less sparse, leading to degraded performance of CSC-modified_$\ell_1$. Fig. 9 further plots the FAN under different values of $\alpha$.

The results together Fig. 8 again confirm that, with a moderate-to-high SNR, the proposed CSC-modified_$\ell_1$ outperforms CS-$\ell_1$, even using less sensor nodes.

## VI. CONCLUSIONS

For CS-based WSNs, this paper proposed a sensor censoring scheme to reduce data transmission energy, while enabling the acquisition of reliable local sensor inference for global sparse signal reconstruction. To the best of our knowledge, our work is an original contribution to the study of sensor censoring for CS networked sensing from a physical-layer perspective. The proposed approach was built on sparse sensing, which allowed each sensor node to infer certain partial knowledge about the unknown signal support. This naturally led to a ternary censoring protocol and an associated optimal design formulation. The optimal censoring rule was analytically derived by using the standard Lagrange multiplier techniques; moreover, by leveraging certain monotonic property of the LR function, a low-complexity implementation based on simple bi-section search was then obtained. To aid global signal reconstruction under the proposed censoring framework, a modified $\ell_1$-minimization based algorithm, which further exploits sparsity of the received hard decision vector at the FC, was proposed. Analytic RIP-based performance guarantees were also given. Computer simulations showed that, in the medium-to-high SNR region, the proposed scheme achieved better signal reconstruction performance using fewer sensors as compared to the conventional CS-based processing without censoring. It would be interesting to extend the proposed censoring protocol to an inter-sensor collaboration setting, e.g., [46]; in general, if the sensors are allowed to communicate among themselves, more reliable local decision outcomes will be obtained. In our scheme, such improved reliability can further enhance the global signal recovery performance, since the proposed signal reconstruction method relies on promoting the sparsity of $\Phi_{\mathcal{I}_{-1}}\mathbf{s}$ (recall that, the more reliable the local decisions are, the sparser the vector $\Phi_{\mathcal{I}_{-1}}\mathbf{s}$ will be). Nevertheless, development of a sparsity-aware collaborative sensor censoring scheme is quite challenging, because the closed-form formulae for the local decision probability and the global signal estimation distortion are very difficult to be derived. Furthermore, with inter-sensor collaboration, the energy consumption cost needs to be redefined; hence the problem formulation and design criterion for sensing censoring require to be reformulated in a very different way. Developing a new sparse-sensing based cooperative censoring scheme needs extensive and careful study. Other future work will also take into account the effect of communication link errors; in addition, the development of censoring schemes under a Bayesian formulation is currently under investigation.

# APPENDIX

*A. Proof of Theorem 3.1*

By using the Lagrange multiplier, solving the optimization problem ($P1$) is equivalent to minimizing the Lagrangian function $\mathcal{L}_i$, given by

$$\mathcal{L}_i = P_M^{(i)} + \lambda_1^{(i)}\left(P_F^{(i)} - \beta_i\right) + \lambda_2^{(i)}\left(P_C^{(i)} - \alpha_i\right)$$
$$= \int_{\mathcal{R}_0^{(i)}} p(z_i \mid \mathcal{T} \cap \mathcal{A}_i \neq \varnothing)dz_i + \lambda_1^{(i)}\int_{\mathcal{R}_1^{(i)}} p(z_i \mid \mathcal{T} \cap \mathcal{A}_i = \varnothing)dz_i$$
$$+ \lambda_2^{(i)}\left[\pi_0^{(i)}\int_{\mathcal{R}_2^{(i)}} p(z_i \mid \mathcal{T} \cap \mathcal{A}_i = \varnothing)dz_i + \pi_1^{(i)}\int_{\mathcal{R}_2^{(i)}} p(z_i \mid \mathcal{T} \cap \mathcal{A}_i \neq \varnothing)dz_i\right] - \lambda_1^{(i)}\beta_i - \lambda_2^{(i)}\alpha_i, \quad \text{(A.1)}$$

where $\pi_0^{(i)}$ and $\pi_1^{(i)}$ are defined in (3.4). Since $\mathcal{R}_0^{(i)}$, $\mathcal{R}_1^{(i)}$ and $\mathcal{R}_2^{(i)}$ are disjoint and $\mathbb{R} = \mathcal{R}_0^{(i)} \cup \mathcal{R}_1^{(i)} \cup \mathcal{R}_2^{(i)}$, we have

$$\int_{\mathcal{R}_2^{(i)}} p(z_i \mid \mathcal{T} \cap \mathcal{A}_i = \varnothing)dz_i = 1 - \int_{\mathcal{R}_0^{(i)}} p(z_i \mid \mathcal{T} \cap \mathcal{A}_i = \varnothing)dz_i - \int_{\mathcal{R}_1^{(i)}} p(z_i \mid \mathcal{T} \cap \mathcal{A}_i = \varnothing)dz_i, \quad \text{(A.2)}$$

and

$$\int_{\mathcal{R}_2^{(i)}} p(z_i \mid \mathcal{T} \cap \mathcal{A}_i \neq \varnothing)dz_i = 1 - \int_{\mathcal{R}_0^{(i)}} p(z_i \mid \mathcal{T} \cap \mathcal{A}_i \neq \varnothing)dz_i - \int_{\mathcal{R}_1^{(i)}} p(z_i \mid \mathcal{T} \cap \mathcal{A}_i \neq \varnothing)dz_i. \quad \text{(A.3)}$$

Hence, by using (A.2) and (A.3) together with some manipulations, the Lagrangian function $\mathcal{L}_i$ in (A.1) can be further expressed as

$$\mathcal{L}_i = \int_{\mathcal{R}_0^{(i)}} \underbrace{\left\{(1 - \lambda_2^{(i)}\pi_1^{(i)})p(z_i \mid \mathcal{T} \cap \mathcal{A}_i \neq \varnothing) - \lambda_2^{(i)}\pi_0^{(i)}p(z_i \mid \mathcal{T} \cap \mathcal{A}_i = \varnothing)\right\}}_{\triangleq I_0^{(i)}(z_i)} dz_i$$
$$+ \int_{\mathcal{R}_1^{(i)}} \underbrace{\left\{(\lambda_1^{(i)} - \lambda_2^{(i)}\pi_0^{(i)})p(z_i \mid \mathcal{T} \cap \mathcal{A}_i = \varnothing) - \lambda_2^{(i)}\pi_1^{(i)}p(z_i \mid \mathcal{T} \cap \mathcal{A}_i \neq \varnothing)\right\}}_{\triangleq I_1^{(i)}(z_i)} dz_i$$
$$- \lambda_1^{(i)}\beta_i - \lambda_2^{(i)}\alpha_i + \lambda_2^{(i)}(\pi_0^{(i)} + \pi_1^{(i)}). \quad \text{(A.4)}$$

(A.4) asserts that, to minimize $\mathcal{L}_i$, the two integral terms should be kept as small as possible. This accordingly enforces the following rule for selecting $\mathcal{R}_0^{(i)}$, $\mathcal{R}_1^{(i)}$ and $\mathcal{R}_2^{(i)}$:

$$\text{If } I_0^{(i)}(z_i) > 0 \text{ and } I_1^{(i)}(z_i) > 0, \quad z_i \text{ should belong to } \mathcal{R}_2^{(i)},$$
$$\text{If } I_0^{(i)}(z_i) > 0 \text{ and } I_1^{(i)}(z_i) < 0, \quad z_i \text{ should belong to } \mathcal{R}_1^{(i)},$$
$$\text{If } I_0^{(i)}(z_i) < 0 \text{ and } I_1^{(i)}(z_i) > 0, \quad z_i \text{ should belong to } \mathcal{R}_0^{(i)},$$
$$\text{If } I_0^{(i)}(z_i) < I_1^{(i)}(z_i) < 0, \qquad z_i \text{ should belong to } \mathcal{R}_0^{(i)},$$
$$\text{If } I_1^{(i)}(z_i) < I_0^{(i)}(z_i) < 0, \qquad z_i \text{ should belong to } \mathcal{R}_1^{(i)}. \qquad (A.5)$$

To proceed, we need the following lemma.

**Lemma A.1:** We claim that $1 - \lambda_2^{(i)} \pi_1^{(i)} > 0$.

*[Proof]:* Done at the end of the appendix. $\qquad\square$

Using the definition of LR in (3.5) and Lemma A.1, the rule of assignment (A.5) can be rewritten as

$$\text{If } \frac{\lambda_2^{(i)} \pi_0^{(i)}}{1 - \lambda_2^{(i)} \pi_1^{(i)}} < L(z_i) < \frac{\lambda_1^{(i)} - \lambda_2^{(i)} \pi_0^{(i)}}{\lambda_2^{(i)} \pi_1^{(i)}}, \qquad z_i \text{ should belong to } \mathcal{R}_2^{(i)},$$

$$\text{If } L(z_i) > \max\left\{ \frac{\lambda_2^{(i)} \pi_0^{(i)}}{1 - \lambda_2^{(i)} \pi_1^{(i)}}, \frac{\lambda_1^{(i)} - \lambda_2^{(i)} \pi_0^{(i)}}{\lambda_2^{(i)} \pi_1^{(i)}} \right\}, \qquad z_i \text{ should belong to } \mathcal{R}_1^{(i)},$$

$$\text{If } L(z_i) < \min\left\{ \frac{\lambda_2^{(i)} \pi_0^{(i)}}{1 - \lambda_2^{(i)} \pi_1^{(i)}}, \frac{\lambda_1^{(i)} - \lambda_2^{(i)} \pi_0^{(i)}}{\lambda_2^{(i)} \pi_1^{(i)}} \right\}, \qquad z_i \text{ should belong to } \mathcal{R}_0^{(i)},$$

$$\text{If } \frac{\lambda_1^{(i)} - \lambda_2^{(i)} \pi_0^{(i)}}{\lambda_2^{(i)} \pi_1^{(i)}} < L(z_i) < \frac{\lambda_2^{(i)} \pi_0^{(i)}}{1 - \lambda_2^{(i)} \pi_1^{(i)}} \text{ and } L(z_i) < \lambda_1^{(i)}, \quad z_i \text{ should belong to } \mathcal{R}_0^{(i)},$$

$$\text{If } \frac{\lambda_1^{(i)} - \lambda_2^{(i)} \pi_0^{(i)}}{\lambda_2^{(i)} \pi_1^{(i)}} < L(z_i) < \frac{\lambda_2^{(i)} \pi_0^{(i)}}{1 - \lambda_2^{(i)} \pi_1^{(i)}} \text{ and } L(z_i) > \lambda_1^{(i)}, \quad z_i \text{ should belong to } \mathcal{R}_1^{(i)}. \qquad (A.6)$$

Thanks to the following inequality (with proof given at the end of the appendix)

$$\frac{\lambda_1^{(i)} - \lambda_2^{(i)} \pi_0^{(i)}}{\lambda_2^{(i)} \pi_1^{(i)}} > \frac{\lambda_2^{(i)} \pi_0^{(i)}}{1 - \lambda_2^{(i)} \pi_1^{(i)}}, \qquad (A.7)$$

the last two conditions in (A.6) can be precluded, and the results directly lead to (3.6). It thus remains to prove Lemma A.1 and (A.7).

*[Proof of Lemma A.1]:* The proof is done by contradiction. Assume otherwise $1 - \lambda_2^{(i)} \pi_1^{(i)} \leq 0$. Then, with some manipulations, the rule of assignment (A.5) can be rewritten as

$$\text{If } L(z_i) < \frac{\lambda_1^{(i)} - \lambda_2^{(i)}\pi_0^{(i)}}{\lambda_2^{(i)}\pi_1^{(i)}}, \quad z_i \text{ should belong to } \mathcal{R}_0^{(i)},$$

$$\text{If } \frac{\lambda_1^{(i)} - \lambda_2^{(i)}\pi_0^{(i)}}{\lambda_2^{(i)}\pi_1^{(i)}} < L(z_i) < \lambda_1^{(i)}, \quad z_i \text{ should belong to } \mathcal{R}_0^{(i)},$$

$$\text{If } L(z_i) > \max\left(\frac{\lambda_1^{(i)} - \lambda_2^{(i)}\pi_0^{(i)}}{\lambda_2^{(i)}\pi_1^{(i)}}, \lambda_1^{(i)}\right), \quad z_i \text{ should belong to } \mathcal{R}_1^{(i)}. \tag{A.8}$$

Suppose that $\max\left(\frac{\lambda_1^{(i)} - \lambda_2^{(i)}\pi_0^{(i)}}{\lambda_2^{(i)}\pi_1^{(i)}}, \lambda_1^{(i)}\right) = \lambda_1^{(i)}$. Then (A.8) can be further expressed as

$$\text{If } L(z_i) < \lambda_1^{(i)}, \quad z_i \text{ should belong to } \mathcal{R}_0^{(i)},$$
$$\text{If } L(z_i) > \lambda_1^{(i)}, \quad z_i \text{ should belong to } \mathcal{R}_1^{(i)}. \tag{A.9}$$

Note that the censoring rule (A.9) activates all the sensor nodes, leading to

$$u_i^*(z_i) = \begin{cases} -1, & \text{if } L(z_i) < \lambda_1^{(i)}; \\ z_i, & \text{if } L(z_i) > \lambda_1^{(i)}. \end{cases} \tag{A.10}$$

We can assume without loss of generality that the false alarm probability is less than one, i.e., $\beta_i < 1$, so that the optimal $\mathcal{R}_0^{(i)*}$ is non-empty. As such, it can be shown that there exists another censoring rule $u_i'$ that achieves a smaller $P_M^{(i)}$, while satisfying the same constraints: This implies $u_i^*$ in (A.10) is not the optimal solution to $(P1)$ and, thus, contradiction is reached.

To obtain such a $u_i'$, we first note that both $L(z_i)$ and $p(z_i)$ are continuous (cf. (E.3)). Pick $z_0 \in \mathcal{R}_0^{(i)*}$ with $\varepsilon \triangleq |L(z_0) - \lambda_1^{(i)}| > 0$. The continuity of $L(z_i)$ guarantees the existence of a $\delta > 0$ such that $|L(z_i) - L(z_0)| < \varepsilon$ whenever $|z_i - z_0| < \delta$. Hence, there exists $0 < \delta' < \delta$ such that $[z_0 - \delta', z_0 + \delta'] \subset \mathcal{R}_0^{(i)*}$ and $\int_{z_0-\delta'}^{z_0+\delta'} p(z_i)dz_i \le \alpha_i$. Choose $u_i'$ to be

$$u_i'(z_i) = \begin{cases} -1, & \text{if } L(z_i) < \lambda_1^{(i)} \text{ and } z_i \notin [z_0 - \delta', z_0 + \delta']; \\ z_i, & \text{if } L(z_i) > \lambda_1^{(i)}; \\ 0, & \text{if } z_i \in [z_0 - \delta', z_0 + \delta']. \end{cases} \tag{A.11}$$

Note that, in (A.11), the local decision region associated with the outcome $u_i'(z_i) = z_i$, say $\mathcal{R}_1^{(i)'}$, is the same as $\mathcal{R}_1^{(i)*}$ of $u_i^*$; hence, $u_i'$ also achieves the same false-alarm probability. Moreover, from (A.11), we note that the no-send region $\mathcal{R}_2^{(i)'} = [z_0 - \delta', z_0 + \delta']$ is non-empty: this implies $\mathcal{R}_0^{(i)'}$ must be a proper subset of $\mathcal{R}_0^{(i)*}$. As a result, $u_i'$ achieves a smaller $P_M^{(i)}$ as compared to $u_i^*$. Therefore, $\lambda_1^{(i)}$ cannot be greater than $\frac{\lambda_1^{(i)} - \lambda_2^{(i)}\pi_0^{(i)}}{\lambda_2^{(i)}\pi_1^{(i)}}$. Next we consider the other case $\max\left(\frac{\lambda_1^{(i)} - \lambda_2^{(i)}\pi_0^{(i)}}{\lambda_2^{(i)}\pi_1^{(i)}}, \lambda_1^{(i)}\right) = \frac{\lambda_1^{(i)} - \lambda_2^{(i)}\pi_0^{(i)}}{\lambda_2^{(i)}\pi_1^{(i)}}$. Then the second condition in (A.8) can be precluded, and thus (A.8) is reduced to

$$\text{If } L(z_i) < \frac{\lambda_1^{(i)} - \lambda_2^{(i)}\pi_0^{(i)}}{\lambda_2^{(i)}\pi_1^{(i)}}, \quad z_i \text{ should belong to } \mathcal{R}_0^{(i)},$$

$$\text{If } L(z_i) > \frac{\lambda_1^{(i)} - \lambda_2^{(i)}\pi_0^{(i)}}{\lambda_2^{(i)}\pi_1^{(i)}}, \quad z_i \text{ should belong to } \mathcal{R}_1^{(i)}. \quad (A.12)$$

Clearly, if we replace $\frac{\lambda_1^{(i)} - \lambda_2^{(i)}\pi_0^{(i)}}{\lambda_2^{(i)}\pi_1^{(i)}}$ by $\lambda_1^{(i)}$ in (A.12), then (A.12) and (A.9) are the same rule.

Therefore, following the similar procedures as in the above proof, we can also show that the assumption $\lambda_1^{(i)} < \frac{\lambda_1^{(i)} - \lambda_2^{(i)}\pi_0^{(i)}}{\lambda_2^{(i)}\pi_1^{(i)}}$ is not true. Hence, the proof is completed. □

*[Proof of (A.7)]:* (A.7) can be proved by using a contradiction-based argument, as what has been done in the proof of Lemma A.1. Assume otherwise $\frac{\lambda_1^{(i)} - \lambda_2^{(i)}\pi_0^{(i)}}{\lambda_2^{(i)}\pi_1^{(i)}} \leq \frac{\lambda_2^{(i)}\pi_0^{(i)}}{1 - \lambda_2^{(i)}\pi_1^{(i)}}$. Then the first condition in (A.6) can be precluded, and (A.6) is reduced to

$$\text{If } L(z_i) > \frac{\lambda_2^{(i)}\pi_0^{(i)}}{1 - \lambda_2^{(i)}\pi_1^{(i)}}, \quad z_i \text{ should belong to } \mathcal{R}_1^{(i)},$$

$$\text{If } L(z_i) < \frac{\lambda_1^{(i)} - \lambda_2^{(i)}\pi_0^{(i)}}{\lambda_2^{(i)}\pi_1^{(i)}}, \quad z_i \text{ should belong to } \mathcal{R}_0^{(i)},$$

$$\text{If } \frac{\lambda_1^{(i)} - \lambda_2^{(i)}\pi_0^{(i)}}{\lambda_2^{(i)}\pi_1^{(i)}} < L(z_i) < \lambda_1^{(i)} < \frac{\lambda_2^{(i)}\pi_0^{(i)}}{1 - \lambda_2^{(i)}\pi_1^{(i)}}, \quad z_i \text{ should belong to } \mathcal{R}_0^{(i)},$$

$$\text{If } \frac{\lambda_1^{(i)} - \lambda_2^{(i)}\pi_0^{(i)}}{\lambda_2^{(i)}\pi_1^{(i)}} < \lambda_1^{(i)} < L(z_i) < \frac{\lambda_2^{(i)}\pi_0^{(i)}}{1 - \lambda_2^{(i)}\pi_1^{(i)}}, \quad z_i \text{ should belong to } \mathcal{R}_1^{(i)}. \quad (A.13)$$

Consequently, with some manipulations, the censoring rule is

$$u_i^*(z_i) = \begin{cases} -1, \text{ if } L(z_i) < \lambda_1^{(i)}; \\ z_i, \text{ if } L(z_i) > \lambda_1^{(i)}, \end{cases} \quad (A.14)$$

which is the same as (A.10). However, as shown in the proof of Lemma A.1, $u_i^*$ is not the optimal censoring rule since we can develop a "better" censoring rule $u_i'$ defined as in (A.11). Hence, the proof is completed by contradiction. □

## B. Proof of Lemma 3.2

Based on Assumption 3, $\mathbf{\Phi}_i^T \mathbf{v}_i$ is Gaussian with zero mean and variance $E\{|\mathbf{\Phi}_i^T \mathbf{v}_i|^2\} = \sigma_v^2 \mathbf{\Phi}_i^T \mathbf{\Phi}_i = K_c \sigma_v^2$. Moreover, conditioned on $|\mathcal{T} \cap \mathcal{A}_i| = j$, $1 \leq j \leq K$, the signal component $\mathbf{\Phi}_i^T \mathbf{s}$ is also Gaussian with zero mean and variance

$$E\{|\mathbf{\Phi}_i^T \mathbf{s}|^2 \| \mathcal{T} \cap \mathcal{A}_i| = j\} \overset{(a)}{=} \sum_{\mathcal{T} \in \mathbf{\Omega}_K} \sum_{q \in \mathcal{T} \cap \mathcal{A}_i} E\{(\Phi_{iq} s_q)^2\} \Pr\{\mathcal{T} \| \mathcal{T} \cap \mathcal{A}_i| = j\} \overset{(b)}{=} j\sigma_s^2, \quad (B.1)$$

where (a) follows from Assumption 2 and (b) is true since $\Phi_{ij} \in \{\pm 1\}$ and $|\mathcal{T} \cap \mathcal{A}_i| = j$. Hence, the conditional probability density functions $p(z_i | \mathcal{T} \cap \mathcal{A}_i \neq \varnothing)$ and $p(z_i | \mathcal{T} \cap \mathcal{A}_i = \varnothing)$ can be respectively obtained as follows

$$p(z_i | \mathcal{T} \cap \mathcal{A}_i \neq \varnothing) = \sum_{j=1}^{K} p(\mathbf{\Phi}_i^T \mathbf{s} + \mathbf{\Phi}_i^T \mathbf{v}_i \| \mathcal{T} \cap \mathcal{A}_i| = j) \Pr\{|\mathcal{T} \cap \mathcal{A}_i| = j | \mathcal{T} \cap \mathcal{A}_i \neq \varnothing\}$$

$$= \sum_{j=1}^{K} \frac{1}{\sqrt{2\pi(j\sigma_s^2 + K_c \sigma_v^2)}} \exp\left(-\frac{z_i^2}{2(j\sigma_s^2 + K_c \sigma_v^2)}\right) \Pr\{|\mathcal{T} \cap \mathcal{A}_i| = j | \mathcal{T} \cap \mathcal{A}_i \neq \varnothing\}, \quad (B.2)$$

and

$$p(z_i | \mathcal{T} \cap \mathcal{A}_i = \varnothing) = p(\mathbf{\Phi}_i^T \mathbf{v}_i) = \frac{1}{\sqrt{2\pi K_c \sigma_v^2}} \exp\left(-\frac{z_i^2}{2 K_c \sigma_v^2}\right), \quad (B.3)$$

Based on Assumption 1, we obtain $\Pr\{|\mathcal{T} \cap \mathcal{A}_i| = j | \mathcal{T} \cap \mathcal{A}_i \neq \varnothing\} = C_j^{K_c} C_{K-j}^{N-K_c} / (\sum_{j=1}^{K} C_j^{K_c} C_{K-j}^{N-K_c})$ for all $j$. The proof of Lemma 3.2 is thus completed. □

## C. Proof of Lemma 3.3

Since $P_j > 0$, $\sqrt{\frac{K_c \sigma_v^2}{(j\sigma_s^2 + K_c \sigma_v^2)}} > 0$, and $\exp\left(\frac{j\sigma_s^2 z_i^2}{2K_c \sigma_v^2 (j\sigma_s^2 + K_c \sigma_v^2)}\right)$ is a monotonically increasing function of $z_i^2$ for all $j$, the function $L$ is a monotonically increasing function of $z_i^2$. Hence, $L_0 = L|_{\mathbb{R}^+ \cup \{0\}}$ is monotonically increasing and thus is one-to-one. The proof is thus completed. □

## D. Proof of Theorem 3.4

By definition of the restriction $L_0(\cdot)$ in Lemma 3.3, the optimal censoring rule $u_i^*$ in (3.6) can be further rewritten as

$$u_i^*(z_i) = \begin{cases} -1, & \text{if } L_0(|z_i|) < \dfrac{\lambda_2^{(i)}\pi_0}{1-\lambda_2^{(i)}\pi_1}; \\[2mm] z_i, & \text{if } L_0(|z_i|) > \dfrac{\lambda_1^{(i)}-\lambda_2^{(i)}\pi_0}{\lambda_2^{(i)}\pi_1}; \\[2mm] 0, & \text{if } \dfrac{\lambda_2^{(i)}\pi_0}{1-\lambda_2^{(i)}\pi_1} < L_0(|z_i|) < \dfrac{\lambda_1^{(i)}-\lambda_2^{(i)}\pi_0}{\lambda_2^{(i)}\pi_1}, \end{cases} \quad (D.1)$$

where $\pi_0 = C_K^{N-K_c}/C_K^N$ and $\pi_1 = \sum_{j=1}^{K} C_j^{K_c} C_{K-j}^{N-K_c}/C_K^N$. Since the inverse function of $L_0(\cdot)$ exists, the assertion immediately follows from (D.1). $\square$

*E. Proof of Theorem 3.5*

We note from (3.12) that there is one-to-one correspondence between the two pairs of variables $(\tau_1^{(i)}, \tau_2^{(i)})$ and $(\lambda_1^{(i)}, \lambda_2^{(i)})$. Hence, an equivalent formulation of Problem ($P1$) is

$$\begin{aligned} \min_{\tau_1^{(i)},\tau_2^{(i)}} \quad & \Pr(|z_i| < \tau_1^{(i)} \mid \mathcal{T} \cap \mathcal{A}_i \neq \varnothing) \\ \text{s.t.} \quad & \Pr\{\tau_1^{(i)} < |z_i| < \tau_2^{(i)}\} \leq \alpha_i, \\ & \Pr(|z_i| > \tau_2^{(i)} \mid \mathcal{T} \cap \mathcal{A}_i = \varnothing) \leq \beta_i. \end{aligned} \quad (E.1)$$

Using (3.7), the objective function in (E.1) is given by

$$\Pr(|z_i| < \tau_1^{(i)} \mid \mathcal{T} \cap \mathcal{A}_i \neq \varnothing) = \sum_{j=1}^{K} P_j \left(1 - 2Q\left(\frac{\tau_1^{(i)}}{\sqrt{j\sigma_s^2 + K_c\sigma_v^2}}\right)\right). \quad (E.2)$$

Note that, since $P_j > 0$ and $Q$ is a monotonically decreasing function of $\tau_1^{(i)}$, minimization of $\Pr(|z_i| < \tau_1^{(i)} \mid \mathcal{T} \cap \mathcal{A}_i \neq \varnothing)$ is equivalent to minimization of $\tau_1^{(i)}$. In addition, with the aid of (3.7) and (3.9), the probability density function $p(z_i)$ can be obtained as follows,

$$\begin{aligned} p(z_i) &= p(z_i \mid \mathcal{T} \cap \mathcal{A}_i = \varnothing)\pi_0^{(i)} + p(z_i \mid \mathcal{T} \cap \mathcal{A}_i \neq \varnothing)\pi_1^{(i)} \\ &= \frac{\pi_0^{(i)}}{\sqrt{2\pi K_c \sigma_v^2}} \exp\left(-\frac{z_i^2}{2K_c\sigma_v^2}\right) + \sum_{j=1}^{K} P_j \frac{\pi_1^{(i)}}{\sqrt{2\pi(j\sigma_s^2 + K_c\sigma_v^2)}} \exp\left(-\frac{z_i^2}{2(j\sigma_s^2 + K_c\sigma_v^2)}\right). \end{aligned} \quad (E.3)$$

Therefore, the probabilities $\Pr\{\tau_1^{(i)} < |z_i| < \tau_2^{(i)}\}$ and $\Pr(|z_i| > \tau_2^{(i)} \mid \mathcal{T} \cap \mathcal{A}_i = \varnothing)$ can be respectively expressed as

$$\Pr\{\tau_1^{(i)} < |z_i| < \tau_2^{(i)}\} = g(\tau_1^{(i)}) - g(\tau_2^{(i)}), \quad (E.4)$$

and

$$\Pr(|z_i| > \tau_2^{(i)} \mid \mathcal{T} \cap \mathcal{A}_i = \varnothing) = 2Q\left(\frac{\tau_2^{(i)}}{\sqrt{K_c}\sigma_v}\right). \quad (E.5)$$

Hence, through some straightforward manipulations, the optimization problem (E.1) can be simplified as

$$\min_{\tau_1^{(i)},\tau_2^{(i)}} \tau_1^{(i)}, \text{ s.t. } \tau_2^{(i)} \geq \sqrt{K_c}\sigma_v Q^{-1}\left(\frac{\beta_i}{2}\right), \text{ and } g(\tau_1^{(i)}) \leq \alpha_i + g(\tau_2^{(i)}). \tag{E.6}$$

Given the first constraint in (E.6), the maximal value of $g(\tau_2^{(i)})$ is thus $g\left(\sqrt{K_c}\sigma_v Q^{-1}\left(\frac{\beta_i}{2}\right)\right)$: This implies the maximal value of $g(\tau_1^{(i)})$ is $\alpha_i + g\left(\sqrt{K_c}\sigma_v Q^{-1}\left(\frac{\beta_i}{2}\right)\right)$. Therefore, the optimization problem (E.6) can be further simplified to

$$\min_{\tau_1^{(i)}} \tau_1^{(i)}, \text{ s.t. } \tau_1^{(i)} \geq g^{-1}\left(\alpha_i + g\left(\sqrt{K_c}\sigma_v Q^{-1}\left(\frac{\beta_i}{2}\right)\right)\right). \tag{E.7}$$

Hence, the optimal $(\tau_1^{(i)},\tau_2^{(i)})$ is given by (3.15) and (3.16). □

*F. Proof of Theorem 3.6*

To derive (3.18), we shall find closed-form formulae for $\Pr\{z_i \in \mathcal{R}_0^{(i)}\}$ and $\Pr\{z_i \in \mathcal{R}_1^{(i)}\}$. Based on Lemma 3.2 and Theorems 3.4 and 3.5, $\Pr\{z_i \in \mathcal{R}_0^{(i)}\}$ and $\Pr\{z_i \in \mathcal{R}_1^{(i)}\}$ can be respectively expressed as

$$\Pr\{z_i \in \mathcal{R}_0^{(i)}\} = \Pr\{|z_i| < \tau_1^{(i)^*}\} = 1 - g(\tau_1^{(i)^*}), \tag{F.1}$$

and

$$\Pr\{z_i \in \mathcal{R}_1^{(i)}\} = \Pr\{|z_i| > \tau_2^{(i)^*}\} = g(\tau_2^{(i)^*}), \tag{F.2}$$

where $g$ is defined in (3.14). With some manipulations, the expected communication cost $C_i$ in (3.17) is given by

$$C_i = C_{0i}\left(1 - g(\tau_1^{(i)^*})\right) + C_{1i}g(\tau_2^{(i)^*}) \overset{(a)}{=} C_{0i}(1 - \alpha_i) + (C_{1i} - C_{0i})g(\tau_2^{(i)^*}), \tag{F.3}$$

where (a) follows from (3.14) and (3.15). Equation (3.18) then follows from (3.16) and (F.3). □

*G. Proof of Theorem 4.1*

By defining $\mathbf{r} \triangleq \mathbf{As}$, problem ($P3$) can be equivalently reformulated as

$$(P4)\ \underset{\mathbf{r}}{Minimize} \|\mathbf{r}\|_1, \text{ subject to } \left\|\mathbf{u}_\mathcal{I} - \mathbf{\Phi}_\mathcal{I}\mathbf{A}^\dagger\mathbf{r}\right\|_2 \leq \varepsilon,\ \left\|\mathbf{r} - \mathbf{A}\mathbf{A}^\dagger\mathbf{r}\right\|_2 = 0.$$

Since $\frac{1}{|\mathcal{I}|}\mathbf{\Phi}_\mathcal{I}\mathbf{A}^\dagger$ satisfies RIP of order $2K$ with $\delta_{2K} < \sqrt{2} - 1$, by Lemma 1.6 in [6] and with some manipulations, the reconstruction error $\mathbf{e} \triangleq \hat{\mathbf{r}} - \mathbf{r}$ obeys

$$\|\mathbf{e}\|_2 \leq \frac{2\left[1-(1-\sqrt{2})\delta_{2K}\right]\sigma_K(\mathbf{r})_1}{\sqrt{K}\left[1-(1+\sqrt{2})\delta_{2K}\right]} + \frac{2}{1-(1+\sqrt{2})\delta_{2K}} \frac{\left|\left\langle \frac{1}{|\mathcal{I}|}\boldsymbol{\Phi}_\mathcal{I}\mathbf{A}^\dagger\mathbf{e}_\Lambda, \frac{1}{|\mathcal{I}|}\boldsymbol{\Phi}_\mathcal{I}\mathbf{A}^\dagger\mathbf{e}\right\rangle\right|}{\|\mathbf{e}_\Lambda\|_2} \quad (G.1)$$

with probability at least $1-\exp\left(-c_3|\mathcal{I}|\varepsilon'\right)$, where $c_3 > 0$ is a constant $\varepsilon' = \left[(\varepsilon/\sqrt{K_c|\mathcal{I}|}\sigma_v) - 1\right]^2$, and $\mathbf{e}_\Lambda \in \mathbb{R}^{|\mathcal{I}_{-1}|+N}$ is obtained by retaining the entries of $\mathbf{e}$ indexed by $\Lambda = \Lambda_0 \cup \Lambda_1$, in which $\Lambda_0$ and $\Lambda_1$ are the index subsets corresponding to the $K$ largest entries of $\mathbf{r}$ and $\mathbf{e}_{\Lambda_0^c}$, respectively. Note that

$$\left|\left\langle \tfrac{1}{|\mathcal{I}|}\boldsymbol{\Phi}_\mathcal{I}\mathbf{A}^\dagger\mathbf{e}_\Lambda, \tfrac{1}{|\mathcal{I}|}\boldsymbol{\Phi}_\mathcal{I}\mathbf{A}^\dagger\mathbf{e}\right\rangle\right| \leq \left\|\tfrac{1}{|\mathcal{I}|}\boldsymbol{\Phi}_\mathcal{I}\mathbf{A}^\dagger\mathbf{e}_\Lambda\right\|_2 \left\|\tfrac{1}{|\mathcal{I}|}\boldsymbol{\Phi}_\mathcal{I}\mathbf{A}^\dagger\mathbf{e}\right\|_2 \overset{(a)}{\leq} \sqrt{1+\delta_{2K}} \|\mathbf{e}_\Lambda\|_2 \left\|\tfrac{1}{|\mathcal{I}|}\boldsymbol{\Phi}_\mathcal{I}\mathbf{A}^\dagger\mathbf{e}\right\|_2$$
$$\overset{(b)}{\leq} 2\frac{\varepsilon}{|\mathcal{I}|}\sqrt{1+\delta_{2K}}\|\mathbf{e}_\Lambda\|_2, \quad (G.2)$$

where (a) follows from the RIP and (b) follows from the constraint in (P4) and the Cauchy-Schwarz inequality. Hence, by means of (G.2), (G.1) can be further expressed as

$$\|\hat{\mathbf{r}} - \mathbf{r}\|_2 \leq \frac{2\left[1-(1-\sqrt{2})\delta_{2K}\right]\sigma_K(\mathbf{r})_1}{\sqrt{K}\left[1-(1+\sqrt{2})\delta_{2K}\right]} + \frac{4\varepsilon\sqrt{1+\delta_{2K}}}{|\mathcal{I}|\left[1-(1+\sqrt{2})\delta_{2K}\right]}. \quad (G.3)$$

Since the optimal solution to $\hat{\mathbf{s}}$ to (P3) is given by $\hat{\mathbf{s}} = \mathbf{A}^\dagger \hat{\mathbf{r}}$, where $\hat{\mathbf{r}}$ is the optimal solution to (P4), it follows from (G.3) that

$$\|\hat{\mathbf{s}} - \mathbf{s}\|_2 \leq \frac{\|\hat{\mathbf{r}} - \mathbf{r}\|_2}{\sigma_{\min}(\mathbf{A})} \leq \frac{2\left[1-(1-\sqrt{2})\delta_{2K}\right]}{\sigma_{\min}(\mathbf{A})\left[1-(1+\sqrt{2})\delta_{2K}\right]} \frac{\sigma_K(\mathbf{A}\mathbf{s})_1}{\sqrt{K}} + \frac{4\sqrt{1+\delta_{2K}}}{\sigma_{\min}(\mathbf{A})\left[1-(1+\sqrt{2})\delta_{2K}\right]}\frac{\varepsilon}{|\mathcal{I}|} \quad (G.4)$$

with probability at least $1-\exp\left(-c_3|\mathcal{I}|\varepsilon'\right)$. The proof is thus completed. $\square$

*H. Proof of Theorem 4.2*

The proof basically consists of two parts. We will first prove that $\frac{1}{\sqrt{\rho}}\boldsymbol{\Phi}_i$ is an isotropic sub-Gaussian random vector [6, Chap. 5]. Then, with the aid of Theorem 2.1 in [47] and some manipulations, the proof is completed. The first part is established directly by the following lemma.

***Lemma H.1:*** Let $\mathbf{q} \in \mathbb{R}^N$ be a $K_c$-sparse vector with support $\mathcal{T}_\mathbf{q} \subset \{1, \cdots, N\}$ uniformly drawn from the collection $\boldsymbol{\Omega}_{K_c} \triangleq \{\mathcal{T}_1, \cdots, \mathcal{T}_{C_{K_c}^N}\}$ of all $C_{K_c}^N = N!/[K_c!(N-K_c)!]$ possible sparsity patterns. The nonzero entries of $\mathbf{q}$ are assumed to be independent symmetric Bernoulli random variables, i.e., $q_i \in \{\pm 1\}$ with $\Pr\{q_i = 1\} = \Pr\{q_i = -1\} = 1/2$ for $i \in \mathcal{T}_\mathbf{q}$. Then $\frac{1}{\sqrt{\rho}}\mathbf{q}$ is an isotropic sub-Gaussian random vector with constant $\alpha = \overline{c}/\sqrt{\rho}$, where $\overline{c} > 0$ is a constant and $\rho = K_c/N$.

*[Proof of Lemma H.1]:* First, for each $1 \leq i \leq C_{K_c}^N$, straightforward manipulations show that the conditional expectation $E\{\mathbf{q}\mathbf{q}^T | \mathcal{T}_\mathbf{q} = \mathcal{T}_i\} = \mathbf{C}_i$, where $\mathbf{C}_i \in \mathbb{R}^{N \times N}$ is a diagonal matrix with $[\mathbf{C}_i]_{jj} = 1$ if $j \in \mathcal{T}_i$ and $[\mathbf{C}_i]_{jj} = 0$ when otherwise. Then, the *second moment matrix* of $\frac{1}{\sqrt{\rho}}\mathbf{q}$ can be obtained as follows,

$$E\{\tfrac{1}{\sqrt{\rho}}\mathbf{q}\tfrac{1}{\sqrt{\rho}}\mathbf{q}^T\} = \frac{1}{\rho}E\{\mathbf{q}\mathbf{q}^T\} = \frac{1}{\rho}\sum_{i=1}^{C_{K_c}^N} E\{\mathbf{q}\mathbf{q}^T | \mathcal{T}_\mathbf{q} = \mathcal{T}_i\}\Pr(\mathcal{T}_\mathbf{q} = \mathcal{T}_i) = \frac{1}{\rho C_{K_c}^N}\sum_{i=1}^{C_{K_c}^N} \mathbf{C}_i = \mathbf{I}_N. \tag{H.1}$$

Hence, by definition, the random vector $\frac{1}{\sqrt{\rho}}\mathbf{q}$ is isotropic. To prove $\frac{1}{\sqrt{\rho}}\mathbf{q}$ is a sub-Gaussian random vector, we need to check that, for every $\mathbf{a} \in \mathbb{R}^N$, the inner product $<\frac{1}{\sqrt{\rho}}\mathbf{q},\mathbf{a}>$ is sub-Gaussian random variable. To see this, let $t \geq 0$ and then we have

$$\begin{aligned}
\Pr\left(\left|<\tfrac{1}{\sqrt{\rho}}\mathbf{q},\mathbf{a}>\right|>t\right) &= \sum_{i=1}^{C_{K_c}^N}\Pr\left(\left|<\tfrac{1}{\sqrt{\rho}}\mathbf{q},\mathbf{a}>\right|>t\Big|\mathcal{T}_\mathbf{q}=\mathcal{T}_i\right)\Pr(\mathcal{T}_\mathbf{q}=\mathcal{T}_i) \\
&= \sum_{i=1}^{C_{K_c}^N}\Pr\left(\left|\sum_{j\in\mathcal{T}_i}\tfrac{1}{\sqrt{\rho}}q_j a_j\right|>t\right)\Pr(\mathcal{T}_\mathbf{q}=\mathcal{T}_i) \\
&\stackrel{(a)}{\leq} \sum_{i=1}^{C_{K_c}^N}\left[e\cdot\exp\left(-\frac{\rho c t^2}{\|\mathbf{a}_{\mathcal{T}_i}\|_2^2}\right)\right]\Pr(\mathcal{T}_\mathbf{q}=\mathcal{T}_i) \\
&\leq \sum_{i=1}^{C_{K_c}^N}\left[e\cdot\exp\left(-\frac{\rho c t^2}{\|\mathbf{a}\|_2^2}\right)\right]\Pr(\mathcal{T}_\mathbf{q}=\mathcal{T}_i) = e\cdot\exp\left(-\frac{\rho c t^2}{\|\mathbf{a}\|_2^2}\right),
\end{aligned} \tag{H.2}$$

where (a) holds due to the fact that $q_j$'s are independent symmetric Bernoulli random variables for all $j \in \mathcal{T}_i$ and thus, by Proposition 5.10 in [6, Chap. 5], the inequality $\Pr\left(\left|\sum_{j\in\mathcal{T}_i}\tfrac{1}{\sqrt{\rho}}q_j a_j\right|\geq t\right) \leq e\exp\left(-\frac{\rho c t^2}{\|\mathbf{a}_{\mathcal{T}_i}\|_2^2}\right)$ is valid, where $\mathbf{a}_{\mathcal{T}_i} \in \mathbb{R}^{K_c}$ is obtained by keeping the entries of $\mathbf{a}$ indexed by $\mathcal{T}_i$ and $c > 0$ is an absolute constant. Inequality (H.2) shows that the random vector $\frac{1}{\sqrt{\rho}}\mathbf{q}$ is sub-Gaussian random vector and the corresponding sub-Gaussian norm is bounded above by $\bar{c}/\sqrt{\rho}$, where $\bar{c} > 0$. Therefore, $\frac{1}{\sqrt{\rho}}\mathbf{q}$ is an isotropic sub-Gaussian random vector in $\mathbb{R}^N$ with constant $\bar{c}/\sqrt{\rho}$. □

The proof of the second part is mainly based on the next two lemmas. To proceed, we define $\mathcal{V}(\mathcal{X}_K) \triangleq \{\mathbf{y} \in \mathbb{R}^N | \mathbf{y} = \mathbf{A}^\dagger \mathbf{x}, \text{ where } \mathbf{x} \in \mathcal{X}_K\}$ to be the restriction of $\mathbf{A}^\dagger$ to the set $\mathcal{X}_K \triangleq \{\mathbf{x} \in \mathbb{R}^{|\mathcal{I}_{-1}|+N} | \|\mathbf{A}^\dagger\mathbf{x}\|_2 = 1, \|\mathbf{x}\|_0 = K\}$, and $\ell_*(\mathcal{V}(\mathcal{X}_K)) \triangleq E\left[\sup_{\mathbf{v}\in\mathcal{V}(\mathcal{X}_K)}|\langle\mathbf{v},\mathbf{u}\rangle|\right]$ to be the complexity measure of the set $\mathcal{V}(\mathcal{X}_K)$, where $\mathbf{u} \sim \mathcal{N}(\mathbf{0},\mathbf{I})$ is a Gaussian random vector [47].

***Lemma H.2 [47, Thm. 2.1]:*** For each $\theta \in (0,1)$, if $|\mathcal{I}|$ satisfies

$$|\mathcal{I}| > \frac{c'\bar{c}^4}{\rho^2\theta^2}\ell_*(\mathcal{V}(\mathcal{X}_K))^2, \tag{H.3}$$

in which $\ell_*(\mathcal{V}(\mathcal{X}_K))$ is the complexity measure of the set $\mathcal{V}(\mathcal{X}_K)$, then with probability at least $1 - \exp\left(-c''\rho^2\theta^2|\mathcal{I}|/\bar{c}^4\right)$, the inequality

$$1 - \theta \leq \frac{\|\mathbf{\Phi}_\mathcal{I}\mathbf{v}\|_2^2}{\rho|\mathcal{I}|} \leq 1 + \theta \tag{H.4}$$

holds for all $\mathbf{v} \in \mathcal{V}(\mathcal{X}_K)$, where $c', c'' > 0$ are constants. □

***Lemma H.3:*** The complexity measure of the set $\mathcal{V}(\mathcal{X}_K)$ is bounded above as follows,

$$\ell_*(\mathcal{V}(\mathcal{X}_K)) \leq \frac{6s_{\max}(\mathbf{A}^\dagger, K)}{s_{\min}(\mathbf{A}^\dagger, K)}\sqrt{K\log\left(\frac{5e(|\mathcal{I}_{-1}| + N)}{K}\right)}. \tag{H.5}$$

*[Proof]:* Done at the end of this appendix. □

We assume that

$$|\mathcal{I}| > \frac{c_1 K s_{\max}^2(\mathbf{A}^\dagger, K)}{\rho^2\theta^2 s_{\min}^2(\mathbf{A}^\dagger, K)}\log\left(\frac{5e(|\mathcal{I}_{-1}| + N)}{K}\right), \tag{H.6}$$

where $c_1 \triangleq 36c'\bar{c}^4$ is a positive absolute constant. Notably, with the aid of (H.5) it is easy to see (H.6) is a sufficient condition on $|\mathcal{I}|$ that guarantees (H.4) to hold. By definition of $\mathcal{V}(\mathcal{X}_K)$, (H.4) reads

$$(1-\theta)\rho\|\mathbf{A}^\dagger\mathbf{v}\|_2^2 \leq \left\|\tfrac{1}{|\mathcal{I}|}\mathbf{\Phi}_\mathcal{I}\mathbf{A}^\dagger\mathbf{v}\right\|_2^2 \leq (1+\theta)\rho\|\mathbf{A}^\dagger\mathbf{v}\|_2^2, \tag{H.7}$$

for all $\mathbf{v} \in \{\mathbf{x} \in \mathbb{R}^{|\mathcal{I}_{-1}|+N} \mid \|\mathbf{x}\|_0 = K\}$. Furthermore, by definitions of $s_{\min}(\mathbf{A}^\dagger, K)$ and $s_{\max}(\mathbf{A}^\dagger, K)$, (H.7) can be further expressed as

$$(1-\theta)\rho s_{\min}^2(\mathbf{A}^\dagger, K)\|\mathbf{v}\|_2^2 \leq \left\|\tfrac{1}{|\mathcal{I}|}\mathbf{\Phi}_\mathcal{I}\mathbf{A}^\dagger\mathbf{v}\right\|_2^2 \leq (1+\theta)\rho s_{\max}^2(\mathbf{A}^\dagger, K)\|\mathbf{v}\|_2^2. \tag{H.8}$$

With some straightforward manipulations, it can be verified that, for $\delta_K \in \left(1 - \rho s_{\min}^2(\mathbf{A}^\dagger, K), 1\right)$ and with $\theta = 1 - \left(\frac{1-\delta_K}{\rho s_{\min}^2(\mathbf{A}^\dagger, K)}\right)$, we have

$$1 - \delta_K = (1-\theta)\rho s_{\min}^2(\mathbf{A}^\dagger, K) \text{ and } 1 + \delta_K \geq (1+\theta)\rho s_{\max}^2(\mathbf{A}^\dagger, K). \tag{H.9}$$

Combining (H.8) with (H.9) yields

$$(1-\delta_K)\|\mathbf{v}\|_2^2 \leq \left\|\tfrac{1}{|\mathcal{I}|}\mathbf{\Phi}_\mathcal{I}\mathbf{A}^\dagger\mathbf{v}\right\|_2^2 \leq (1+\delta_K)\|\mathbf{v}\|_2^2. \tag{H.10}$$

Using (H.6) and Lemma H.2, it can be concluded that (H.10) holds with probability exceeding

$1 - \exp\left(-c_2\rho^2\theta^2(\delta_K)|\mathcal{I}|\right)$ for any $K$-sparse vector $\mathbf{v}$, where $c_2 \triangleq c''/\overline{c}^4 > 0$ is a constant and $\theta(\delta_K) = 1 - \left(\dfrac{1-\delta_K}{\rho s_{\min}^2(\mathbf{A}^\dagger, K)}\right)$. The proof is thus completed. $\square$

*[Proof of Lemma H.3]:* Let $\mathcal{V}_K^{|\mathcal{I}_{-1}|+N} \triangleq \{\mathbf{v} \in \mathcal{S}^{|\mathcal{I}_{-1}|+N-1} \mid \|\mathbf{v}\|_0 \leq K\}$ be the subset of the unit Euclidean sphere $\mathcal{S}^{|\mathcal{I}_{-1}|+N-1}$ in $\mathbb{R}^{|\mathcal{I}_{-1}|+N}$ and $\mathbf{u} \sim \mathcal{N}(\mathbf{0}, \mathbf{I}_N)$ be a Gaussian random vector. Then we have

$$\ell_*(\mathcal{V}(\mathcal{X}_K)) = E\left[\sup_{\mathbf{v}\in\mathcal{V}(\mathcal{X}_K)} |\langle \mathbf{v}, \mathbf{u}\rangle|\right] = E\left[\sup_{\mathbf{x}\in\mathcal{X}_K} \|\mathbf{x}\|_2 \left|\left\langle \mathbf{A}^\dagger \tfrac{\mathbf{x}}{\|\mathbf{x}\|_2}, \mathbf{u}\right\rangle\right|\right]$$

$$\leq \sup_{\mathbf{x}\in\mathcal{X}_K} \|\mathbf{x}\|_2 \, E\left[\sup_{\mathbf{r}\in\mathcal{V}_K^{|\mathcal{I}_{-1}|+N}} \left|\langle \mathbf{A}^\dagger \mathbf{r}, \mathbf{u}\rangle\right|\right]$$

$$\leq \frac{1}{s_{\min}(\mathbf{A}^\dagger, K)} E\left[\sup_{\mathbf{r}\in\mathcal{V}_K^{|\mathcal{I}_{-1}|+N}} \left|\langle \mathbf{A}^\dagger \mathbf{r}, \mathbf{u}\rangle\right|\right]$$

$$\overset{(a)}{\leq} \frac{6 s_{\max}(\mathbf{A}^\dagger, K)}{s_{\min}(\mathbf{A}^\dagger, K)} \sqrt{K \log\left(\frac{5e(|\mathcal{I}_{-1}|+N)}{K}\right)},$$

where (a) follows from Lemma B.6 in [48]. Hence the proof is completed. $\square$

## REFERENCES

[1] R. G. Baraniuk, "Compressive sensing," *IEEE Signal Processing Magazine*, vol. 24, no. 4, pp, 118-124, July 2007.
[2] E. J. Candes and M. B. Wakin, "An introduction to compressive sampling," *IEEE Signal Processing Magazine*, vol. 25, no. 2, pp, 21-30, March 2008.
[3] E. J. Candes, J. Romberg, and T. Tao, "Stable signal recovery from incomplete and inaccurate measurements," *Comm. Pure Appl. Math.*, vol. 59, no. 8, pp. 1207-1223, 2006.
[4] E. J. Candes, "The restricted isometry property and its implications in compressive sensing," *Comptes Rendus Mathematique*, vol. 346, issue 9-10, pp. 589-592, May 2008.
[5] S. Foucart and H. Rauhut, *A Mathematical Introduction to Compressive Sensing*, Birkhauser, 2010.
[6] Y. C. Eldar, and G. Kutyniok, *Compressed Sensing: Theory and Applications*, Cambridge University Press, 2011.
[7] J. D. Haupt, W. U. Bajwa, M. Rabbat, and R. D. Nowak, "Compressed sensing for networked data," *IEEE Signal Processing Magazines*, vol. 25, no. 2, pp. 92-101, March 2008.
[8] W. U. Bajwa, J. D. Haupt, A. M. Sayeed, and R. D. Nowak, "Joint source-channel communications for distributed estimation in sensor networks," *IEEE Trans. Information Theory*, vol. 53, no. 10, pp. 3629-3653, Oct. 2007.
[9] A. Y. Yang, M. Gastpar, R. Bajcsy, and S .S. Sastry, "Distributed sensor perception via sparse representation," *Proceedings of the IEEE,* vol. 98, no. 6, pp. 1077-1088, June 2010.
[10] Y. Zheng, N. Cao, T. Wimalajeewa, and P. K. Varshney, "Compressive sensing based probabilistic sensor management for target tracking in wireless sensor networks," *IEEE Trans. Signal Processing*, vol. 63, no. 22, pp. 6049-6060, Nov. 2015.
[11] T. Wimalajeewa, and P. K. Varshney, "Wireless compressive sensing over fading channels with distributed sparse random projections," *IEEE Trans. Signal and Information Processing over Networks*, vol. 1, no. 1, pp. 33-44, March 2015.
[12] G. Quer, R. Masiero, G. Pillonetto, M. Rossi and M. Zorzi, "Sensing, compression, and recovery for WSNs: Sparse signal modeling and monitoring framework," *IEEE Trans. Wireless Communications*, vol. 11, no. 10, pp. 3447–3461, Oct. 2012.
[13] N. M. Gowda, and A. P. Kannu, "Interferer identification in HetNets using compressive sensing framework," *IEEE Trans. Communications*, vol. 61, no. 11, pp. 4780–4787, Nov. 2013.
[14] T. Xue, X. Dong and Y. Shi, "Multiple access and data reconstruction in wireless sensor networks based on compressed sensing," *IEEE Trans. Wireless Communications*, vol. 12, no. 7, pp. 3399–3411, July 2013.
[15] M. Leinonen, M. Codreanu and M. Juntti, "Sequential compressed sensing with progressive signal reconstruction in wireless sensor networks," *IEEE Trans. Wireless Communications*, vol. 14, no. 3, pp. 1622–1635, March 2015.
[16] I. F. Akyildiz and M. C. Vuran, *Wireless Sensor Networks*, John Wiely & Sons, Ltd., 2010.
[17] C. Rago, P. K. Willett, and Y. Bar-Shalom, "Censoring sensors: a low-communication-rate scheme for distributed detection," *IEEE Trans. Aerosp. Electron. Syst.*, vol. 32, no. 2, pp. 554–568, Apr. 1996.